\documentclass{aa} 

\usepackage{graphicx}

\begin{document}
\thesaurus{20 (
           11.01.2; 
           11.03.2; 
	   11.02.1; 
	   11.17.3; 
           13.18.1; 
           13.18.2
)} 
\title{Survey of Instantaneous 1--22~GHz Spectra of 550 Compact
Extragalactic Objects with Declinations from $-30^\circ$ to
$+43^\circ$\thanks{Tables 1 and 5 are available at CDS to \mbox{cdsarc.u-strasbg.fr} \mbox{(130.79.128.5)}
or via \mbox{http://cdsweb.u-strasbg.fr/Abstract.html}}}
\titlerunning{1--22 GHz Spectra of 550 Extragalactic Objects}
\author{Y.Y.~Kovalev\inst{1}
	\and N.A.~Nizhelsky\inst{2}
	\and Yu.A.~Kovalev\inst{1}
	\and A.B.~Berlin\inst{3}
	\and G.V.~Zhekanis\inst{2}
	\and M.G.~Mingaliev\inst{2}
	\and A.V.~Bogdantsov\inst{2} }
\authorrunning{Kovalev et al.}
\offprints{Y.Y.~Kovalev}
\mail{yyk@asc.rssi.ru}
\institute{Astro Space Center of the Lebedev Physical Institute,
           Profsoyuznaya 84/32, Moscow, 119997 Russia
           \and
           Special Astrophysical Observatory,
           Nizhny Arkhyz, Karachaevo--Cherkessia, 357147 Russia
	   \and
	   Special Astrophysical Observatory, St. Petersburg Branch,
	   St. Petersburg, 196140 Russia }
\date{Received March~26; accepted August 13, 1999}
\maketitle

\begin{abstract}
\exhyphenpenalty=10000

     We present observational results for extragalactic radio sources
with milliarcsecond components, obtained with the 600 meter ring radio
telescope RATAN--600 from 1st to 22nd December, 1997. For each source,
a six frequency broad band radio spectrum was obtained by observing
simultaneously with an accuracy up to a minute at 1.4, 2.7, 3.9, 7.7,
13 and 31~cm. The observed list is selected from Preston et al.~(1985)
VLBI survey and contains all the sources in the declinations between
$-30^\circ$ and $+43^\circ$ with a correlated flux density exceeding
0.1~Jy at 13~cm. The sample includes the majority of sources to be
studied in the current VSOP survey and the future RadioAstron Space
VLBI mission.

\end{abstract}

\section{Introduction}

     One of the main characteristics of an extragalactic radio source
is the shape of the broad band spectrum, which provides a considerable
amount of physical information about the object. The extragalactic
radio sources are often separated into different samples on the basis
of the shape of the spectra (e.g. flat, inverted, steep, gigahertz
peaked spectrum sources). This shows the importance of the
multifrequency broad band spectra surveys of compact extragalactic
objects among long term flux variability monitoring programs (e.g.
Aller et al.~\cite{Aller_etal85}, Mitchell et al.~\cite{Mitchell_etal94},
Stevens et al.~\cite{Stevens_etal94}, etc.) and VLBI imaging surveys
(e.g. Kellermann et al.~\cite{Kellermann_etal98}
and references therein).

     Most of the earlier multifrequency spectra results were obtained
by combining measurements carried out quasi-simultaneously (over a
period of one or more months) at radio and shorter wavelengths, using
several telescopes on some samples of tens of objects, selected by
various criteria. For example, we refer to the measurements of
19~active extragalactic sources from 20~cm to 1400~\AA{} by Landau et
al.~(\cite{Landau_etal86}). Valtaoja et al.~(\cite{Valtaoja_etal88})
investigated quiescent spectra for a sample of 27~radio sources observed
at centimeter and millimeter wavelengths. Gear et
al.~(\cite{Gear_etal94}) compared quasi-simultaneous 5--375~GHz spectra
of 22~BL~Lacertae objects with 24~radio-loud, violently variable
quasars. K\"uhr et al.~(\cite{Kuhr_etal81}) compiled radio measurements
at more than two frequencies for 494~sources from the combined NRAO--MPI
5~GHz Strong Source Surveys and the Parkes 2.7~GHz Surveys. Herbig \&
Readhead~(\cite{HerbigReadhead92}) composed non-simultaneous radio data
from 10~MHz to 100~GHz on a complete sample of 256~objects. We also
refer to papers, cited in the O'Dea~(\cite{ODea98}) survey of compact
steep-spectrum and gigahertz peaked-spectrum sources, as well as a
number of other works.

     Some earlier RATAN--600 results on broad band spectra observations
were presented for samples from 8.7~GHz Zelenchuk sky survey by
Amirkhanyan et al.~(\cite{Amir_etal92}) and 87GB survey by Mingaliev \&
Khabrakhmanov~(\cite{MingKhab95}), for strong compact extragalactic
objects by Kovalev et al.~(\cite{Kovalev_etal96}) and weak radio
sources from the RATAN--600 ``Cold" deep sky survey by
Bursov~(\cite{Bursov97}).

     A broad band spectrum of a compact extragalactic radio source
is usually considered to be the sum of spectra of several compact
and extended components in a source structure. The components are
located at various distances from the central nucleus of the object,
and are likely to have resulted from activity within the nucleus.
The nucleus may be a black hole, which converts an accreted ambient
gas to ejected relativistic particles along magnetic fields. Some
of the components can be variable in time. The variability can
misrepresent the true shape of a spectrum if measurements at all or
especially high frequencies are not simultaneous.

     In this work we present observational results of more than five
hundred sources at six frequencies from 1 to 22~GHz using a single
radio telescope. Flux densities at all frequencies are measured
practically instantaneously~-- over a period of a few minutes. This is
the shortest time scale of broad band six frequency measurements for
the largest sample of sources so far, which has been used for a spectra
survey of compact extragalactic radio sources.

     These observations are part of a long-term program of
instantaneous spectra monitoring of compact extragalactic objects
(Kovalev~\cite{Kovalev98}), which have milliarcsecond components and are
studied by VLBI networks. They also give a ground spectra support for
the VSOP survey and a pre-launch spectra study of the objects for the
RadioAstron project. The goal of the long-term program is a mass study
of the spectra and their variability for many hundreds of compact
extragalactic radio sources. It is also among our intentions to find a
relationship between instantaneous multifrequency spectra and the VLBI
radio structure.

\section{Source sample}

     We have selected about 700~sources from the Preston et
al.~(\cite{Preston_etal85}) VLBI survey. These sources have a
correlated flux density $F^\mathrm{corr}_{13}\ge 0.1$~Jy at the
wavelength of 13~cm, and are located north of declination $-30^\circ$.
The northern sector of RATAN--600 restricts this declination range from
$-30^\circ$ to $+43^\circ$. Measurements of the sources located north
of declination $+49^\circ$ were made in 1998 with the southern sector
of RATAN--600 and will be published at a later date.

     The list of 551 sources for northern sector observations is
presented in Table~\ref{list_table}. The columns of this table are as
follows: (1) the IAU name, (2) possible other name, (3) the optical
identification (OI) and (4) the redshift. The OI and redshifts are
taken from Veron--Cetty \& Veron (\cite{VeronVeron98}) or, if not found
there, are taken from the NASA/IPAC Extragalactic database (NED). The
abbreviations used are ``Q" for quasars, ``BL" for BL~Lacertae objects,
``G" for galaxies, ``RS" for radio sources. In the latter case we do
not have OI. An extended version of this table is available in
electronic form at the CDS. This also includes B1950
coordinates taken from Preston et al.~(\cite{Preston_etal85}) and
Morabito et al.~(\cite{Morabito_etal86}), correlated flux densities at
13~cm with errors from Preston et al.~(\cite{Preston_etal85}) and
optical spectrum classifications from Veron--Cetty~\&
Veron~(\cite{VeronVeron98}).

\section{Observations}

     We performed continuous six frequency 1--22 GHz spectra
observations of compact extragalactic sources from 1st to 22nd
December, 1997. We used the 600~meter ring radio telescope RATAN--600
(Korolkov \& Parijskij~\cite{Korolkov_Parijskij79},
Parijskij~\cite{Parijskij93}) at the Russian Academy of Sciences'
Special Astrophysical Observatory, located in Karachaevo-Cherkessia
Republic (Russia) near Nizhny Arkhyz and Zelenchukskaya at the North
Caucasus. The northern sector of the antenna (a part of the main ring
reflector) was used together with the secondary mirror of cabin No.~1.
Six broad band receivers are located at this movable cabin. The cabin
moves along 150~meter long rails in order to be placed in the focus of
the antenna system at different elevations. The main (meridional)
transit method of observation was employed. Accuracy and reliability of
these spectra measurements were essentially higher than earlier
experiments conducted at the RATAN--600 due to the following
improvements made to the receivers, antenna control system and the
procedure of observations.

\setcounter{table}{1}

\begin{table}[t]
\caption{
Parameters of RATAN--600 broad band receivers in 1997,
used in this work
}
\begin{center}
\begin{tabular}{llllllll}
\hline
$\lambda$, & $n_\mathrm{h}$ & $\nu_0$, & $\Delta \nu$, & $T^\mathrm{phys}_\mathrm{LNA}$, & $T_\mathrm{LNA}$, & $T_\mathrm{sys}$, & $\delta T_\mathrm{sys}$,\\
cm         &                & GHz      & GHz           & K                               & K                             & K                 & mK          \\
\hline 
1.4  & 2 & 21.65 & 2.5  & 15  & 23 &  77 & 3.5 \\
2.7  & 2 & 11.2  & 1.4  & 15  & 18 &  70 & 3   \\
3.9  & 2 & 7.70  & 1.0  & 15  & 14 &  62 & 3   \\
7.7  & 1 & 3.90  & 0.6  & 15  &  8 &  37 & 2.5 \\
13   & 1 & 2.30  & 0.4  & 310 & 35 &  95 & 8   \\
31   & 1 & 0.96  & 0.12 & 310 & 21 & 105 & 15  \\
\hline
\end{tabular}
\end{center}
\label{receiv_table}
\end{table}

\begin{table}[t]
\caption{
Measured and estimated beam widths and ratios $r$ for
the RATAN--600 northern sector with the secondary mirror No.~1
for different wavelengths $\lambda$ and elevations $h$
}
\begin{center}
\begin{tabular}{llll}
\hline
$\lambda$, cm & \multicolumn{3}{c}{$\mathrm{HPBW}_\mathrm{RA} \times \mathrm{HPBW}_\mathrm{Dec}$} \\
              & $h=80^\circ$ & $h=40^\circ$ & $h=20^\circ$        \\
              & $r\approx5$  & $r\approx10$ & $r\approx20$        \\
\hline
1.4 & 8\farcs0  $\times$ 35\arcsec & 8\farcs5  $\times$ 1\farcm4  & 13\arcsec $\times$ 4\farcm3   \\
2.7 & 16\arcsec $\times$ 1\farcm3  & 17\arcsec $\times$ 2\farcm8  & 24\arcsec $\times$ 8\farcm0   \\
3.9 & 23\arcsec $\times$ 2\farcm0  & 26\arcsec $\times$ 4\farcm3  & 39\arcsec $\times$ 13\arcmin  \\
7.7 & 48\arcsec $\times$ 4\farcm0  & 53\arcsec $\times$ 8\farcm8  & 1\farcm4  $\times$ 27\arcmin  \\
13  & 1\farcm3  $\times$ 6\farcm5  & 1\farcm3  $\times$ 13\arcmin & 1\farcm8  $\times$ 37\arcmin  \\
31  & 3\farcm2  $\times$ 16\arcmin & 3\farcm3  $\times$ 33\arcmin & 5\farcm0  $\times$ 100\arcmin \\
\hline
\end{tabular}
\end{center}
\label{beam_table}
\end{table}

     We used a new set of broad band receivers at the wavelengths of 1.4,
2.7, 3.9, 7.7, 13 and 31~cm with low noise HEMT amplifiers (LNA),
cooled to a temperature of 15~K at the four shortest wavelengths
(Berlin et al.~\cite{Berlin_etal97}, \cite{Berlin_etal93}). Parameters
of radiometers and antenna beams are given in Tables~\ref{receiv_table},
\ref{beam_table}.

     Table~\ref{receiv_table} lists wavelengths $\lambda$; numbers
of feedhorns $n_\mathrm{h}$; exact central frequencies $\nu_0$; band
widths $\Delta \nu$; physical temperatures of the LNA
$T^\mathrm{phys}_\mathrm{LNA}$; noise temperatures of the LNA
$T_\mathrm{LNA}$; total noise temperatures of systems
$T_\mathrm{sys}$, including the antenna noise at middle elevations;
rms noise temperature sensitivities of the system $\delta
T_\mathrm{sys}$ for a one second integration time.
Dual-feedhorn receivers are beam-switched. Single-feedhorn
receivers have a noise-added, gain-balanced mode of operation.
Linearly polarized systems were available at all frequencies:
horizontal at 7.7~cm and vertical at other wavelengths.

     Table~\ref{beam_table} gives half power beam widths
(HPBW) in right ascensions $\mathrm{HPBW}_\mathrm{RA}$ and
declinations $\mathrm{HPBW}_\mathrm{Dec} = r \,
\mathrm{HPBW}_\mathrm{RA}$, for various elevations. The values of
$\mathrm{HPBW}_\mathrm{RA}$ were obtained from our
measurements. Ignoring the aberration effects, we estimated the
factor $r$ using theoretical simulations of the RATAN--600 beam by
Esepkina et al. (\cite{Esepkina_etal79}) and their experimental testing
by Temirova~(\cite{Temirova83}).
     A map of the knife-like beam of the RATAN--600 northern sector
is known to have different shapes of contours cross-sections
at high and low power levels.
In the absence of aberrations, the shapes can be
described as ordinary elliptical contours (elongated on
declinations) at half power level and higher levels. The contours
are transformed to ``the elongated eight" at lower levels or
to ``a dumb-bell" at 0.1 normalized power
level (see Esepkina et al.~\cite{Esepkina_etal79} for details).

     The full permanent automatic control of 225 elements of the main
ring reflector was achieved using the new control system of the antenna
(Zhekanis~\cite{Zhekanis97}; Golubchin et al.~\cite{Golubchin_etal95}).
Errors in position of each element of the main reflector, if present,
were recorded in order to check the quality of the antenna surface for
each observation. The positioning of cabin No.~1 with the secondary
mirror was measured from one of the eight geodetic reference points,
located every 20~meters along the rails. The accuracy of semi-automatic
positioning of the cabin directed towards the focus was again checked
by us some minutes before each observation. If an error of more than
2~mm with respect to a value given in the schedule was found, it was
corrected.

\begin{figure}[t]
\rotatebox{-90}{
   \resizebox{!}{\hsize}{\includegraphics[trim=0cm 0cm 1cm 0cm]{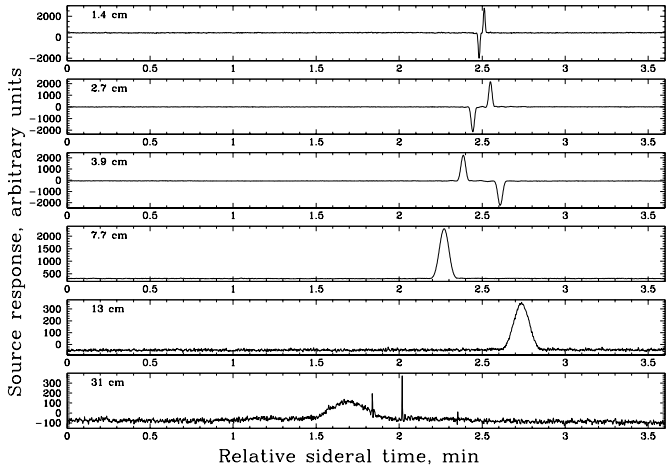}}}
\caption{Example of a full multifrequency scan for \object{4C~39.25},
observed on 16 December 1997 with 0.1 second integration time}
\label{obs_figure}
\end{figure}

     All horns of the radiometers are horizontally located and form a
new configuration, which is an optimal one for decreasing transversal
aberrations. Observations were carried out in the main meridian
(transit mode). As a result, a response to an object is obtained due to
its horizontal scanning by the antenna beam because of the daily
rotation of the Earth (see an example of \object{4C~39.25} full scan on
Fig.~\ref{obs_figure}; the moment of culmination is at $2\fm5$ here).
The total duration of each six frequency observations was usually about
five minutes, and included also two sets of noise temperature
calibration for 30--40 seconds before and after the passing of the
source. The data acquisition system (Chernenkov \&
Tsibulev~\cite{ChernenkovTsibulev95}) controls radiometers and records
the output signals. After each observation, the main ring reflector and
the cabin No.~1 with receivers and the secondary mirror were
repositioned for observation of the next source on a new elevation.

     We have optimized the observational schedule, using new software
(Zhekanis \& Zhekanis~\cite{ZhekanisZhekanis97}). To increase the
reliability of results by final averaging of the spectra, we endeavoured
to include each source in the schedule two or more times during the set
and each flux density calibrator in more than 70\% of the days. The
typical number of sources observed in a 24~hour observing session was
about 80 in the optimized schedule. Several breaks in observations
occurred because of weather conditions (snow-falls or unusually low
temperatures $t<-15^\circ$C), nine hours of technical maintenance per
week, etc. As a result, the total number of successful observations was
about 1450~during 21~days, or about 69~observations per day on average
and 2.6~spectra per source during the set (formal averaging). In 20\%
of the sources the spectra have been measured only once.

\section{Data reduction and calibration}

\begin{figure}[t]
\rotatebox{-90}{
   \resizebox{!}{\hsize}{\includegraphics[trim=0cm 0cm 1cm 0cm]{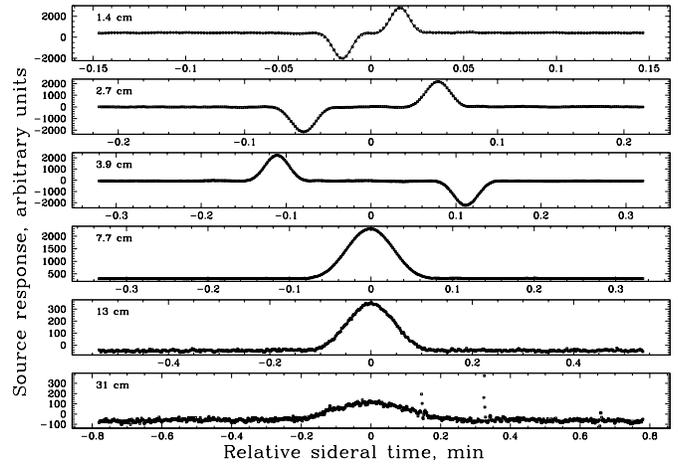}}}
\caption{Example of fitting the simulated beam of RATAN--600
         (lines) to the source response (dots) for \object{4C~39.25} from
	 Fig.~\ref{obs_figure}}
\label{fit_figure}
\end{figure}

     Data reduction has been done using a YURZUF software package,
which had been specially designed for the automatic reduction of the
broad band spectra monitoring observations (Kovalev~\cite{Kovalev98}).
Fitting a beam which is simulated at the source elevation allows us to
compute the amplitude of a source response at each frequency. A
Singular Value Decomposition subroutine from Forsythe et al.
(\cite{Forsythe_etal77}) was applied. Routine functions, designed by
V.R.~Amirkhanyan, were also included via an additional interface as a
subroutine to the YURZUF software to simulate the main antenna beam
together with the secondary lobe. Before the reduction, the quality of
such a fitting has been checked and a simulation of the beam has been
optimized by tuning control parameters using the sample of 30--50
sources which are strong and compact at all frequencies, and
distributed on different elevations (see an example of fitting on
Fig.~\ref{fit_figure}).

\begin{table}[t]
\caption{Parameters of calibration sources:
flux density, Jy, and correction factors $g_\mathrm{ext}$ and $g_\mathrm{pol}$
(from top to bottom)}
\begin{center}
\begin{tabular}{lllllll}
\hline
Source             &        \multicolumn{6}{c}{$\lambda$, cm}      \\
                   & 1.4   & 2.7   & 3.9   & 7.7   & 13    & 31    \\
\hline
\object{0134$+$32} & 1.216 & 2.431 & 3.540 & 6.765 & 10.79 & 21.90 \\
($h$=$79\fdg3$)    & 1.016 & 1.004 & 1.002 & 1.000 & 1.000 & 1.000 \\
                   & 1.032 & 1.046 & 1.039 & 0.959 & 1.004 & 1.002 \\
\object{0237$-$23} & 0.700 & 1.470 & 2.200 & 4.030 & 5.590 & 6.610 \\
($h$=$23\fdg0$)    & 1.000 & 1.000 & 1.000 & 1.000 & 1.000 & 1.000 \\
                   & 0.979 & 0.979 & 0.979 & 1.025 & 0.992 & 0.983 \\
\object{0518$+$16} & 1.135 & 2.008 & 2.700 & 4.413 & 6.221 & 10.28 \\
($h$=$62\fdg8$)    & 1.000 & 1.000 & 1.000 & 1.000 & 1.000 & 1.000 \\
                   & 0.954 & 0.956 & 0.923 & 1.104 & 0.930 & 0.964 \\
\object{0624$-$05} & 1.400 & 2.764 & 4.156 & 8.112 & 12.80 & 24.10 \\
($h$=$40\fdg3$)    & 1.034 & 1.009 & 1.004 & 1.001 & 1.000 & 1.000 \\
                   & 1.027 & 1.020 & 1.021 & 1.025 & 0.942 & 1.017 \\
\object{1328$+$30} & 2.563 & 4.244 & 5.529 & 8.576 & 11.50 & 17.20 \\
($h$=$76\fdg7$)    & 1.016 & 1.004 & 1.002 & 1.000 & 1.000 & 1.000 \\
                   & 1.053 & 0.948 & 0.957 & 1.049 & 0.954 & 0.954 \\
\object{2037$+$42} & \dots & \dots & \dots & 17.40 & 12.10 & 5.000 \\
($h$=$88\fdg5$)    & \dots & \dots & \dots & 1.099 & 1.077 & 1.013 \\
                   & \dots & \dots & \dots & 1.000 & 1.000 & 1.000 \\
\object{2105$+$42} & 5.330 & 5.940 & 6.100 & 5.050 & 2.850 & \dots \\
($h$=$88\fdg4$)    & 1.297 & 1.078 & 1.034 & 1.008 & 1.003 & \dots \\
                   & 0.992 & 0.992 & 0.999 & 1.000 & 1.000 & \dots \\
\hline
\end{tabular}
\end{center}
\label{calibr_table}
\end{table}

\begin{figure}[t]
\hskip 30pt
\resizebox{\hsize}{!}{
   \includegraphics[trim=0cm 0cm 0cm 0.7cm]{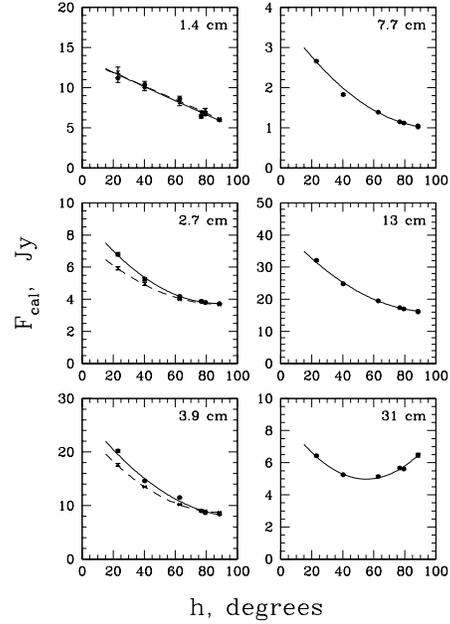}}
\caption{The flux density calibration factor $F_\mathrm{cal}$
versus elevation $h$ at all wavelengths. Solid and
dashed lines at 1.4, 2.7 and 3.9~cm represent $F_\mathrm{cal}$ for each
horn separately. All data (crosses and points with errors), shown
for 6--7 calibrators, were averaged during the observational set.
Seven calibrators are shown at 7.7 and 13~cm, but the data for two
calibrators at $h$=$88\fdg4$ and $h$=$88\fdg5$ are plotted in the same spot
}
\label{cal_figure}
\end{figure}

     The following seven flux density calibrators were applied
to obtain the calibration curve in the scale of Baars et al.
(\cite{Baars_etal77}): \object{0134+32}, \object{0237$-$23},
\object{0518+16}, \object{0624$-$05}, \object{1328+30},
\object{2037+42} (for calibration at 7.7, 13 and 31~cm only),
\object{2105+42} (excluding 31~cm calibration).
They were recommended by Baars et al.~(\cite{Baars_etal77}),
excluding \object{0237$-$23} which is the traditional
RATAN--600 flux density calibrator at low elevations.
Measurements of some calibrators were corrected, where
necessary, on angular size and linear polarization, following the
data, summarized in Ott et al.~(\cite{Ott_etal94}) and Tabara \&
Inoue~(\cite{TabaraInoue80}) respectively. Response to an extended
calibrator was simulated as a two-dimensional convolution of the beam
and brightness distribution in the published model of a calibrator.
The best fit to the observed response was found by optimization of
the angular size of an extended calibrator at each frequency. The
correction factor due to an angular extension $g_\mathrm{ext}$ was
calculated numerically by integrating over the solid angle of the
optimized brightness distribution and the convolution of the
distribution with the beam. Following Ott et al. (\cite
{Ott_etal94}), we applied Gaussian profiles of the brightness
distribution over right ascension and declination for
\object{0134+32}, \object{0625$-$05}, \object{1328+30},
\object{2037+42} and the elliptical disk model for
\object{2105+42}, additionally making an axial ratio to be
equal to the measured one in Masson (\cite{Masson89}). The
correction factor due to linear polarization $g_\mathrm{pol}$ of
the calibrators was calculated in the standard way (Kuzmin \&
Salomonovich~\cite{Kuzmin_Salomonovich64}, Kraus~\cite{Kraus66}) as
$g_\mathrm{pol}=1/[1+p\cos (2\varphi)]$, where $p$ is the linear
polarization degree and $\varphi$ is the angle between polarization
planes of a source and the antenna. The corrected amplitude of
the response to a calibrator is calculated as the observed one,
multiplied by the factors $g_\mathrm{ext}$ and $g_\mathrm{pol}$.

     The flux densities of the calibrators, in Janskies, the
factors $g_\mathrm{ext}$ and $g_\mathrm{pol}$ are summarized in
Table~\ref{calibr_table} for each source (at the elevation $h$)
at each wavelength from top to bottom respectively. The flux densities
were calculated from polynomial approximations
(Taylor~\cite{Taylor99}) of the VLA measurements (relative to
the spectrum of \object{3C~295}) for \object{0134+32},
\object{0518+16}, \object{1328+30}; from spline and polynomial
approximations of the data by Ott et al. (\cite{Ott_etal94}) for
\object{0624$-$05}, \object{2037+42}, \object{2105+42}
(in relation to the spectra of \object{3C~295} and
\object{3C~286}), and from the polynomial approximation (K\"uhr et
al.~\cite{Kuhr_etal79}) of the spectrum for
\object{0237$-$23}. For \object{0134+32} and \object{1328+30} at
31~cm, we give preference to the flux densities extrapolated from
the approximations of Ott et al.~(\cite{Ott_etal94}). Taking into
account all available data, we also used two following extrapolated
values: 0.70~Jy for \object{0237$-$23} and 1.4~Jy for
\object{0624$-$05} at the wavelength of 1.4~cm. With these
extrapolations, we have obtained reasonable results.

     Amplitude measurement of a source in flux density units
has been done in relation to an amplitude of a flux density
calibrator by comparing both with the amplitude of a stable
signal from a noise generator, using standard methods.
In our observations the elevations of seven flux density
calibrators are fixed. Because of this fact we computed a
regression curve to obtain dependence of the flux density
calibration factor $F_\mathrm{cal}$ on the elevation for each horn
on each frequency (Fig.~\ref{cal_figure}). $F_\mathrm{cal}$ is equal
to the amplitude of the noise generator signal, calibrated in flux
density units. In fact, the obtained calibration curves
$F_\mathrm{cal}(h)$ show the dependence of the mean measured flux density
for a source on the elevation $h$, if its
antenna temperature is equal to that of the noise generator
signal $T_\mathrm{ns}$ (Kovalev~\cite{Kovalev98}):
$F_\mathrm{cal}(h)=2kT_\mathrm{ns}/A^\mathrm{aa}_\mathrm{eff}(h)$, where
$A^\mathrm{aa}_\mathrm{eff}(h)= A_\mathrm{eff}(h)\, q^\mathrm{ab}(h)\,
q^\mathrm{atm}(h)$,
$A_\mathrm{eff}(h)$ is the effective area of the antenna in the focus,
$q^\mathrm{ab}(h)$ -- the factor of aberration
(due to transversal shifts of a feedhorn from the focus),
$q^\mathrm{atm}(h)$ -- the atmosphere attenuation factor,
$k$ -- the Boltsman's constant.
We did not make any additional atmospheric correction during
the set (the altitude of RATAN--600 site is 970~m above sea-level).

     The total relative rms error of each individual flux density
measurement $\sigma/F$ is estimated from the following
relation (Kovalev \cite{Kovalev98}):
$$
  \left(\frac{\sigma}{F}\right)_{\nu}^2 =
  \left(\frac{\sigma_\mathrm{s}}{A_\mathrm{s}}\right)_{\nu}^2 +
  \left(\frac{\sigma_\mathrm{ns}}{A_\mathrm{ns}}\right)_{\nu}^2 +
  \left(\frac{\sigma_\mathrm{cal}}{F_\mathrm{cal}}\right)_{\nu}^2 +
  \left(\sigma^\mathrm{r}_\mathrm{{scale}}\right)_{\nu}^2 \, ,
$$
where the first term inside the brackets on the right hand is
the relative error of the amplitude $A_\mathrm{s}$ of a
source (after fitting the simulated response to the observed one),
the second~-- the relative error of the amplitude
$A_\mathrm{ns}$ of a response to the noise generator signal,
the third~-- the relative error of our flux density calibration
$F_\mathrm{cal}$, averaged on the set,
and the last~-- the relative error
$\sigma^\mathrm{r}_\mathrm{scale}$ of the absolute flux density
scale.
Usually, the last term is excluded from presented errors,
but we show it in the relation to emphasize its importance,
because different calibrators may be used in various works.

     The total error $\sigma$ is calculated,
excluding only the $\sigma^\mathrm{r}_\mathrm{scale}$ error,
which is estimated by Baars et al.~(\cite{Baars_etal77}), Ott
et al.~(\cite{Ott_etal94}) and Taylor~(\cite{Taylor99}) as
about 10\% at 1.4~cm and 3--5\% at other wavelengths. It is
better to increase $\sigma^\mathrm{r}_\mathrm{scale}$ to
10--15\% at 1.4~cm for the sources with declinations less than
$-5^\circ$ because of the above mentioned extrapolation of the
flux density values for \object{0624$-$05} and
\object{0237$-$23}. Errors
$(\sigma_\mathrm{cal}/F_\mathrm{cal})$ of the calibration
depend on elevation and are formally less than 2.6, 0.7, 1.4,
1.1, 0.7 and 0.9 \% at 1.4, 2.7, 3.9, 7.7, 13 and 31~cm
respectively (the errors are averaged here over two horns at
1.4--3.9~cm, Fig.~\ref{cal_figure}).

     Mean values are always calculated as the mean weighted values,
if several measurements have been made, with corrections by the
Student's factor to increase the reliability to the standard value
0.683 for one sigma error. The dispersion of frequent measurements
of calibrators (and, consequently, $\sigma_\mathrm{cal}$ and
$\sigma$) as well as calculated errors of mean spectra measurements
represent also random instrumental instabilities and a variability
due to atmosphere conditions during the set.

     Systematic errors caused by various reasons including
calibration are known to be often the main real errors. We have
compared our results with published observations of
other authors to check the residual systematic errors, using several
tens of strong objects distributed on elevations with constant or
slightly variable broad band spectra. The agreement is found to be
quite good within the total accuracy of the data.

\section{Results}

     Table~5 with the results of observations is available in the
electronic form only at the CDS via anonymous ftp. It includes the flux
density data (with one sigma errors without scale errors
$\sigma^\mathrm{r}_\mathrm{scale}$) at 1.4, 2.7, 3.9, 7.7, 13 and 31~cm
for 546 of 551 objects of the source list from Table~\ref{list_table}.
The results of instantaneous observations of the spectra are shown in
Fig.~4. Averaged instantaneous spectra are given in Table~5 and Fig.~4,
if two or more observations of a source have been done.

     The sources \object{0156$-$14}, \object{1635$-$03} were not
observed in December, 1997. We have pointed the antenna to the
coordinates of \object{0611+13} several times, but we have not detected
an emission from the source at any frequency (nothing is present at the
coordinates of \object{0611+13} in NED too). The object \object{3C~274}
is resolved by RATAN--600 at all observed frequencies; multifrequency
response to \object{3C~111} has a double maximum. We have excluded the
data for these sources from final results. 

     Absence of data for some sources at some frequencies is a
result, in general, of data exclusion for the following reasons:
possible confusion in declinations (especially at low elevations) or
partial resolution of a source at some frequencies (e.g.
\object{3C~154} at 1.4, 2.7, 3.9, 7.7~cm), a source is too weak to
be measured reliably (e.g. \object{III~Zw~2} at 31~cm), a
strong influence of man-made interferences (frequently at 31~cm,
sometimes at 13~cm), strong interferences from a stationary placed
satellite at 2.7 and 7.7~cm (in declinations between $-10^\circ$
and $0^\circ$). Nevertheless, in some cases the data was not
excluded in spite of the increase in errors caused usually by
interferences.

     We believe that shapes of the instantaneous spectra presented can
be explained by continuous activity of the nuclei of the objects in
accordance with the basic hypothesis of a source with {\it two
dominating} general components (compact and extended), following
Kovalev et al.~(\cite{Kovalev_etal96}), Kovalev \&
Kovalev~(\cite{Kov_Kov96}). The detailed analysis of the data is
deferred to later papers.

\begin{acknowledgements}

     We would like to thank the RATAN--600 staff for technical support
of the observational process. Some of our problems in achieving the
twenty four hour observations were resolved through the valuable
assistance of Ira Morozova. We are obliged to Harry Ter\"asranta for
providing us with the unpublished data of 22~GHz Mets\"ahovi
observations to check our calibration at this frequency, to Greg Taylor
for observing the object \object{0237$-$23} at our request at the VLA
in 1998 and 1999 in order to study known high frequency discrepancy of
the source spectrum, and to Vladimir Amirkhanyan for kindly making his
routine on simulating the RATAN--600 beam available at our disposal. We
thank Tanya Downs and an anonymous referee for carefully reading the
manuscript and for valuable comments. YYK and YAK are grateful to the
administration and employees of the observatory for their hospitality
during the visit for carrying out the observations. This work has been
partly supported by the Russian State Program ``Astronomy"
(grant~1.2.5.1). YYK acknowledges support from International Soros
Science Educational Program grants a97--2965, a98--1932 and a99--1882.
This research has made use of the NASA/IPAC Extragalactic database
(NED), which is operated by the Jet Propulsion Laboratory, Caltech,
under contract with the National Aeronautic and Space Administration.

\end{acknowledgements}

\setcounter{table}{0}

\clearpage
\begin{table*}[t]
\caption[]{Source sample}
\begin{center}
\begin{tabular}{llll|llll|llll}
\hline
IAU        & Alias       & OI &  $z$     & IAU        & Alias       & OI &  $z$     & IAU        & Alias       & OI & $z$     \\
\hline
0003$+$38  &             & G  &    0.229 & 0159$-$11  & 3C 57       & Q  &    0.669 & 0420$-$01  & OA 129      & Q  &    0.915\\
0003$-$06  & NRAO 5      & BL &    0.347 & 0201$+$11  & OD 101      & Q  &    3.56  & 0421$+$01  & OF 36       & Q  &    2.055\\
0005$-$23  &             & Q  &    1.410 & 0202$+$14  & NRAO 91     & G  &    0.405 & 0422$+$00  & OF 38       & BL &         \\
0005$-$26  & OB $-$210   & G  &          & 0202$+$31  &             & Q  &    1.466 & 0423$+$23  &             & RS &         \\
0007$+$10  & III Zw 2    & G  &    0.090 & 0202$-$17  &             & Q  &    1.740 & 0423$+$05  &             & Q  &    1.333\\
0007$+$17  & 4C 17.04    & Q  &    1.601 & 0211$+$17  &             & Q  &    0.472 & 0425$+$04  & OF 42       & Q  &         \\
0008$-$26  & OB $-$214   & Q  &    1.093 & 0216$+$01  &             & Q  &    1.623 & 0428$+$20  & OF 247      & G  &    0.219\\
0010$+$40  & 4C 40.01    & Q  &    0.256 & 0217$-$18  &             & Q  &          & 0429$+$41  & 3C 119      & Q  &    1.023\\
0011$-$04  &             & Q  &          & 0219$+$42  & 3C 66A      & BL &    0.444 & 0430$+$05  & 3C 120      & G  &    0.033\\
0012$+$31  & 3C 6        & G  &          & 0219$-$16  &             & Q  &    0.698 & 0434$-$18  &             & Q  &    2.702\\
0013$-$00  &             & Q  &    1.574 & 0221$+$06  & 4C 06.11    & Q  &    0.511 & 0440$-$00  & NRAO 190    & Q  &    0.844\\
0019$+$05  & OB 34       & BL &          & 0223$+$34  & 4C 34.07    & Q  &          & 0446$+$11  &             & G  &    1.207\\
0022$+$39  & OA 26       & Q  &    1.946 & 0226$-$03  & 4C $-$03.07 & Q  &    2.066 & 0451$-$28  & OF $-$285   & Q  &    2.560\\
0024$+$34  & OB 338      & G  &    0.333 & 0229$+$13  & 4C 13.14    & Q  &    2.065 & 0454$+$03  & OF 92       & Q  &    1.349\\
0026$+$34  & OB 343      & G  &    0.6   & 0234$+$28  & CTD 20      & Q  &    1.207 & 0454$+$06  & 4C 06.21    & G  &    0.405\\
0027$+$05  &             & Q  &          & 0235$+$16  & OD 160      & BL &    0.940 & 0454$-$23  & OF $-$292   & Q  &    1.003\\
0035$+$23  & CTD 5       & Q  &    2.27  & 0237$-$02  &             & Q  &    1.116 & 0456$+$06  & OF 94       & Q  &    1.08 \\
0035$+$12  & 4C 12.05    & Q  &    1.395 & 0237$+$04  & OD 62       & Q  &    0.978 & 0457$+$02  & OF 97       & Q  &    2.384\\
0035$-$02  & 3C 17       & G  &    0.220 & 0237$-$23  & OD $-$263   & Q  &    2.225 & 0458$-$02  & DA 157      & Q  &    2.286\\
0038$-$02  &             & Q  &    1.176 & 0238$-$08  & NGC 1052    & G  &    0.005 & 0458$+$13  &             & RS &         \\
0047$+$02  & OB 78       & BL &          & 0239$+$10  & OD 166      & Q  &          & 0459$+$06  & OF 99.3     & Q  &    1.106\\
0048$-$09  & OB $-$80    & BL &          & 0240$-$21  & OD $-$267   & Q  &    0.314 & 0459$+$13  &             & BL &         \\
0048$-$07  & OB $-$82    & Q  &    1.975 & 0248$+$43  &             & Q  &    1.310 & 0459$+$25  & 3C 133      & G  &    0.277\\
0054$-$00  &             & Q  &    2.795 & 0250$+$17  &             & Q  &          & 0500$+$01  & OG 3        & Q  &    0.585\\
0055$+$30  & NGC 315     & G  &    0.016 & 0256$+$07  & OD 94.7     & Q  &    0.893 & 0502$+$04  & OG 5        & Q  &    0.954\\
0055$-$05  &             & Q  &          & 0301$+$33  & 4C 33.06    & G  &          & 0507$+$17  &             & G  &    0.416\\
0056$-$00  & DA 32       & Q  &    0.717 & 0306$+$10  & OE 110      & Q  &    0.863 & 0509$+$15  &             & RS &         \\
0106$+$01  & 4C 01.02    & Q  &    2.107 & 0309$+$41  & NRAO 128    & G  &    0.136 & 0511$-$22  & OG $-$220   & Q  &    1.296\\
0108$-$07  & OC $-$14    & Q  &    1.776 & 0312$+$10  & 4C 10.10    & G  &          & 0514$-$16  & OG $-$123   & Q  &    1.270\\
0108$+$38  & OC 314      & G  &    0.668 & 0316$+$16  & CTA 21      & Q  &          & 0518$+$16  & 3C 138      & Q  &    0.759\\
0109$+$22  &             & BL &          & 0316$+$41  & 3C 84       & G  &    0.017 & 0521$-$26  & OG $-$236   & RS &         \\
0110$+$31  & NRAO 62     & Q  &    0.603 & 0317$+$18  & OE 129      & G  &          & 0528$-$25  & OG $-$247   & Q  &    2.765\\
0111$+$02  & UGC 773     & G  &    0.047 & 0319$+$12  & OE 131      & Q  &    2.67  & 0528$+$13  & OG 147      & Q  &    2.07 \\
0112$-$01  &             & Q  &    1.365 & 0322$+$22  &             & RS &          & 0537$-$15  &             & Q  &    0.947\\
0113$-$11  &             & Q  &    0.672 & 0326$+$27  &             & Q  &    1.533 & 0537$-$28  & OG $-$263   & Q  &    3.104\\
0116$+$08  & 4C 08.06    & G  &    0.594 & 0327$-$24  & OE $-$246.3 & Q  &    0.888 & 0552$+$39  & DA 193      & Q  &    2.365\\
0118$-$27  & OC $-$230.4 & BL & $>$0.557 & 0329$-$25  & OE $-$248   & Q  &    2.685 & 0555$-$13  &             & Q  &         \\
0119$+$11  &             & Q  &    0.570 & 0333$+$32  & NRAO 140    & Q  &    1.259 & 0601$+$24  & 4C 24.11    & RS &         \\
0119$+$04  & OC 33       & Q  &    0.637 & 0336$-$01  & CTA 26      & Q  &    0.852 & 0602$+$40  & OH 404.1    & RS &         \\
0119$+$24  &             & Q  &    2.025 & 0338$-$21  & OE $-$263.9 & BL &          & 0605$-$08  & OH $-$10    & Q  &    0.872\\
0122$-$00  &             & Q  &    1.070 & 0340$+$36  & OE 367      & Q  &    1.484 & 0606$-$22  & OH $-$212   & Q  &    1.926\\
0123$+$25  & 4C 25.05    & Q  &    2.364 & 0344$+$19  &             & RS &          & 0607$-$15  & OH $-$112   & Q  &    0.324\\
0127$+$14  & 4C 14.06    & Q  &          & 0346$-$16  &             & Q  &          & 0610$+$26  & 3C 154      & Q  &    0.580\\
0130$-$17  &             & Q  &    1.020 & 0348$-$12  & OE $-$182   & Q  &    1.520 & 0611$+$13$^\mathrm{c}$ & &    &         \\
0133$-$20  & OC $-$255.3 & Q  &    1.141 & 0400$+$25  & CTD 26      & Q  &    2.109 & 0618$-$25  & OH $-$230   & Q  &    1.90 \\
0134$+$32  & 3C 48       & Q  &    0.367 & 0402$+$37  & 4C 37.11    & G  &    0.054 & 0620$+$38  & OH 335      & Q  &    3.469\\
0135$-$24  & OC $-$259   & Q  &    0.831 & 0403$-$13  & OF $-$105   & Q  &    0.571 & 0641$+$39  & OH 368.8    & Q  &    1.266\\
0136$+$17  &             & Q  &    2.716 & 0405$-$12  & OF $-$109   & Q  &    0.574 & 0642$+$21  & 3C 166      & G  &    0.245\\
0138$-$09  & OC $-$65    & BL & $>$0.501 & 0406$+$12  &             & BL &    1.02  & 0650$+$37  &             & Q  &    1.982\\
0142$-$27  & OC $-$270   & Q  &    1.157 & 0406$-$12  & OF $-$111   & Q  &    1.563 & 0653$-$03  & OH $-$90    & Q  &         \\
0144$+$20  &             & RS &          & 0409$+$22  & 3C 108      & Q  &    1.213 & 0711$+$35  & OI 318      & Q  &    1.620\\
0146$+$05  & OC 79       & Q  &    2.345 & 0410$+$11  & 3C 109      & G  &    0.306 & 0722$+$14  & 4C 14.23    & Q  &         \\
0147$+$18  & OC 178      & Q  &          & 0413$-$21  &             & Q  &    0.807 & 0723$-$00  & OI $-$39    & BL &    0.127\\
0148$+$27  &             & Q  &    1.26  & 0414$-$18  &             & Q  &    1.536 & 0727$+$40  & OI 446      & Q  &    2.501\\
0149$+$21  &             & Q  &    1.32  & 0415$+$37$^\mathrm{b}$ & 3C 111 & G & 0.048 & 0727$-$11  &             & RS &         \\
0149$+$33  & OC 383      & Q  &    2.431 & 0420$+$02  &             & BL &          & 0729$+$25  &             & Q  &         \\
0156$-$14$^\mathrm{a}$ & & RS &          & 0420$+$41  & 4C 41.11    & RS &          & 0733$+$30  &             & RS &         \\
\hline
\end{tabular}
\label{list_table}
\end{center}
\end{table*}

\clearpage
\begin{table*}[t]
{\bf Table 1.} continued
\begin{center}
\begin{tabular}{llll|llll|llll}
\hline
IAU        & Alias       & OI &  $z$     & IAU        & Alias       & OI &  $z$     & IAU        & Alias       & OI & $z$     \\
\hline
0733$-$17  &             & RS &          & 0953$+$25  & OK 290      & Q  &    0.712 & 1148$-$17  & OM $-$181   & Q  &    1.751\\
0733$+$26  &             & RS &          & 0955$+$32  & 3C 232      & Q  &    0.530 & 1156$-$22  &             & G  &    0.565\\
0735$+$17  & OI 158      & BL & $>$0.424 & 1004$-$01  &             & Q  &    1.214 & 1156$-$09  & OM $-$94    & RS &         \\
0736$-$06  & OI $-$61    & Q  &    1.901 & 1004$+$14  & OL 108.1    & Q  &    2.707 & 1156$+$29  & 4C 29.45    & Q  &    0.729\\
0736$+$01  & OI 61       & Q  &    0.191 & 1008$-$01  & 4C $-$01.21 & Q  &    0.887 & 1157$-$21  &             & Q  &    0.927\\
0738$+$31  & OI 363      & Q  &    0.630 & 1010$+$35  & OL 318      & Q  &    1.414 & 1200$-$05  & ON $-$1     & Q  &    0.381\\
0738$+$27  &             & RS &          & 1012$+$23  & 4C 23.24    & Q  &    0.565 & 1200$+$04  &             & RS &         \\
0742$+$31  & 4C 31.30    & Q  &    0.462 & 1013$+$20  & OL 224      & Q  &    3.11  & 1202$-$26  &             & Q  &    0.790\\
0742$+$10  & OI 471      & RS &          & 1015$+$35  & OL 326      & Q  &    1.226 & 1204$+$28  & ON 208      & Q  &    2.177\\
0743$-$00  & 4C $-$00.28 & Q  &    0.994 & 1018$+$34  & OL 331      & Q  &    1.400 & 1210$+$13  & 4C 13.46    & Q  &    1.137\\
0743$+$25  &             & RS &          & 1019$+$42  &             & RS &          & 1211$+$33  & ON 319      & Q  &    1.598\\
0743$+$27  &             & RS &          & 1019$+$30  & OL 333      & Q  &    1.319 & 1213$-$17  & ON $-$122   & G  &         \\
0745$+$24  & OI 275      & Q  &    0.409 & 1020$-$10  & OL $-$133   & Q  &    0.197 & 1213$+$35  & 4C 35.28    & Q  &    0.857\\
0748$+$12  & OI 280      & Q  &    0.889 & 1020$+$19  & OL 133      & Q  &    2.136 & 1215$+$30  & ON 325      & BL &    0.237\\
0748$+$33  & OI 380      & Q  &    1.932 & 1020$+$40  & 4C 40.25    & Q  &    1.254 & 1216$-$01  &             & Q  &    0.415\\
0752$-$11  & OI $-$187   & RS &          & 1021$-$00  &             & Q  &    2.552 & 1217$+$02  & ON 29       & Q  &    0.240\\
0754$+$10  & OI 90.4     & BL &    0.66  & 1022$+$19  & 4C 19.34    & Q  &    0.828 & 1218$+$33  & 3C 270.1    & Q  &    1.519\\
0759$+$18  &             & Q  &          & 1030$+$41  &             & Q  &    1.120 & 1218$-$02  & 4C $-$02.53 & G  &    0.665\\
0802$+$21  &             & RS &          & 1030$+$39  &             & Q  &    1.095 & 1219$+$28  & ON 231      & BL &    0.102\\
0805$+$41  &             & Q  &    1.420 & 1032$-$19  &             & Q  &    2.198 & 1219$+$04  & 4C 04.42    & Q  &    0.965\\
0805$+$26  &             & RS &          & 1034$-$05  & OL $-$57    & G  &          & 1222$+$03  & 4C 03.23    & Q  &    0.960\\
0805$-$07  &             & Q  &    1.837 & 1034$-$29  & OL $-$259   & Q  &    0.312 & 1222$+$21  & 4C 21.35    & Q  &    0.435\\
0808$+$01  & OJ 14       & BL &          & 1036$-$15  & OL $-$161   & G  &    0.525 & 1225$+$36  & ON 343      & Q  &    1.975\\
0812$+$36  & OJ 320      & Q  &    1.025 & 1038$+$06  & 4C 06.41    & Q  &    1.265 & 1226$+$02  & 3C 273      & Q  &    0.158\\
0812$+$02  & 4C 02.23    & Q  &    0.402 & 1040$+$12  & 3C 245      & Q  &    1.028 & 1228$+$12$^\mathrm{b}$ & 3C 274 & G & 0.004\\
0814$+$42  & OJ 425      & BL &          & 1042$+$07  &             & G  &    0.698 & 1228$-$11  & ON $-$147   & Q  &    3.528\\
0818$-$12  & OJ $-$131   & BL &          & 1045$-$18  & OL $-$176   & Q  &    0.595 & 1236$+$07  &             & G  &    0.400\\
0820$+$22  & 4C 22.21    & BL &    0.951 & 1046$-$02  & 4C $-$02.43 & RS &          & 1237$-$10  & ON $-$162   & Q  &    0.750\\
0820$+$29  & OJ 234      & Q  &    2.368 & 1054$+$00  & OL 91       & RS &          & 1240$+$38  &             & Q  &    1.316\\
0821$+$39  & 4C 39.23    & Q  &    1.216 & 1055$+$20  & 4C 20.24    & Q  &    1.11  & 1240$-$29  &             & Q  &    1.133\\
0823$+$03  & OJ 38       & BL &    0.506 & 1055$+$01  & OL 93       & Q  &    0.888 & 1243$-$07  & ON $-$73    & Q  &    1.286\\
0827$+$24  & OJ 248      & Q  &    0.941 & 1058$+$39  &             & RS &          & 1244$-$25  &             & Q  &    0.638\\
0829$+$04  & OJ 49       & BL &    0.180 & 1100$+$22  & OM 201      & RS &          & 1252$+$11  & ON 187      & Q  &    0.870\\
0830$+$42  & OJ 451      & Q  &    0.253 & 1101$+$38  & Mark 421    & BL &    0.031 & 1253$-$05  & 3C 279      & Q  &    0.538\\
0834$+$25  & OJ 259      & Q  &    1.122 & 1102$-$24  & OM $-$204   & Q  &    1.666 & 1255$+$32  & ON 393      & RS &         \\
0837$+$03  &             & Q  &    1.57  & 1104$+$16  & 4C 16.30    & Q  &    0.632 & 1256$-$220 & ON $-$293.9 & Q  &    1.306\\
0838$+$13  & 3C 207      & Q  &    0.684 & 1106$+$38  &             & G  &    2.290 & 1256$-$229 &             & Q  &    1.365\\
0839$+$18  &             & Q  &    1.272 & 1109$+$35  &             & RS &          & 1257$+$14  & OW 197      & Q  &         \\
0851$+$07  &             & RS &          & 1110$-$21  & OM $-$218   & RS &          & 1302$-$03  &             & Q  &    1.250\\
0851$+$20  & OJ 287      & BL &    0.306 & 1111$+$14  & OM 118      & Q  &    0.869 & 1302$-$10  & OP $-$106   & Q  &    0.286\\
0854$+$21  &             & RS &          & 1116$+$12  & 4C 12.39    & Q  &    2.118 & 1308$+$32  & OP 313      & BL &    0.997\\
0855$+$14  & 3C 212      & Q  &    1.043 & 1119$+$18  & OM 133      & Q  &    1.040 & 1308$+$14  & OP 114      & Q  &         \\
0859$-$14  & OJ $-$199   & Q  &    1.339 & 1120$-$27  & OM $-$234   & RS &          & 1315$+$34  & OP 326      & Q  &    1.050\\
0900$+$42  & 4C 42.28    & G  &          & 1123$+$26  & CTD 74      & Q  &    2.341 & 1317$-$00  & 4C $-$00.50 & Q  &    0.892\\
0906$+$43  & 3C 216      & Q  &    0.670 & 1124$-$18  &             & Q  &    1.048 & 1318$-$26  &             & Q  &    2.027\\
0906$+$01  & DA 263      & Q  &    1.018 & 1127$-$14  & OM $-$146   & Q  &    1.187 & 1330$+$02  & 3C 287.1    & G  &    0.215\\
0912$+$02  &             & G  &    0.427 & 1128$+$38  &             & Q  &    1.733 & 1331$+$17  & OP 151      & Q  &    2.084\\
0912$+$29  & OK 222      & BL &          & 1128$-$04  & OM $-$48    & G  &    0.266 & 1334$-$12  & OP $-$158.3 & Q  &    0.539\\
0913$+$39  & 4C 38.28    & Q  &    1.269 & 1130$+$00  &             & Q  &          & 1336$-$23  & OP $-$260.5 & Q  &         \\
0915$-$21  &             & Q  &    0.847 & 1136$-$13  & CTS 667     & Q  &    0.554 & 1336$-$26  &             & Q  &    1.51 \\
0922$+$00  & OK 37       & Q  &    1.719 & 1142$+$05  & 4C 05.52    & Q  &    1.342 & 1337$-$03  &             & Q  &         \\
0923$+$39  & 4C 39.25    & Q  &    0.698 & 1142$-$22  & OM $-$271   & Q  &    1.141 & 1345$+$12  & 4C 12.50    & G  &    0.121\\
0925$-$20  &             & Q  &    0.348 & 1143$-$24  & OM $-$272   & Q  &    1.940 & 1347$-$21  & OP $-$279   & RS &         \\
0931$-$11  & OK $-$152   & RS &          & 1143$-$28  & OM $-$273   & Q  &    0.45  & 1348$-$28  &             & Q  &         \\
0938$+$11  &             & Q  &    3.191 & 1144$+$40  &             & Q  &    1.089 & 1349$-$14  & OP $-$182   & RS &         \\
0945$+$40  & 4C 40.24    & Q  &    1.252 & 1145$-$07  & OM $-$76    & Q  &    1.342 & 1351$-$01  &             & Q  &    3.707\\
0952$+$17  & OK 186      & Q  &    1.478 & 1148$-$00  & 4C $-$00.47 & Q  &    1.980 & 1352$-$10  & OP $-$187   & Q  &    0.332\\
\hline
\end{tabular}
\end{center}
\end{table*}

\clearpage
\begin{table*}[t]
{\bf Table 1.} continued
\begin{center}
\begin{tabular}{llll|llll|llll}
\hline
IAU        & Alias       & OI &  $z$     & IAU        & Alias       & OI &  $z$     & IAU        & Alias       & OI & $z$     \\
\hline
1354$-$17  &             & Q  &    3.147 & 1611$+$34  & DA 406      & Q  &    1.401 & 2044$-$16  & OW $-$174   & Q  &    1.932\\
1354$-$15  & OP $-$192   & Q  &    1.89  & 1614$+$05  & OS 23       & Q  &    3.217 & 2047$+$09  &             & RS &         \\
1354$+$19  & 4C 19.44    & Q  &    0.719 & 1615$+$36  & 4C 36.27    & RS &          & 2047$+$03  &             & BL &         \\
1356$+$02  &             & Q  &    1.319 & 1615$+$02  &             & Q  &    1.341 & 2053$-$04  & 4C $-$04.80 & Q  &    1.176\\
1402$-$01  &             & Q  &    2.522 & 1616$+$06  & OS 28       & Q  &    2.088 & 2058$-$29  &             & Q  &    1.492\\
1402$+$04  &             & Q  &    3.211 & 1622$-$25  & OS $-$237.8 & Q  &    0.786 & 2059$+$03  & OW 98       & Q  &    1.015\\
1403$-$08  &             & Q  &    1.763 & 1622$-$29  &             & Q  &    0.815 & 2113$+$29  &             & Q  &    1.514\\
1404$+$28  & OQ 208      & G  &    0.077 & 1624$+$41  & 4C 41.32    & Q  &    2.55  & 2121$+$05  & OX 36       & Q  &    1.941\\
1406$-$07  & OQ $-$10    & Q  &    1.493 & 1625$-$14  &             & Q  &    1.10  & 2126$-$15  & OX $-$146   & Q  &    3.266\\
1406$-$26  &             & Q  &    2.43  & 1633$+$38  & 4C 38.41    & Q  &    1.807 & 2126$-$18  & OX $-$145   & Q  &    0.680\\
1413$+$34  & OQ 323      & RS &          & 1635$-$03$^\mathrm{a}$ & & Q  &    2.856 & 2127$-$09  &             & Q  & $>$0.780\\
1416$+$06  & 3C 298      & Q  &    1.439 & 1638$+$39  & NRAO 512    & Q  &    1.666 & 2128$+$04  & OX 46       & G  &    0.990\\
1427$+$10  & OQ 147      & Q  &    1.71  & 1641$+$39  & 3C 345      & Q  &    0.594 & 2128$-$12  & OX $-$148   & Q  &    0.501\\
1430$-$17  & OQ $-$151   & Q  &    2.331 & 1647$-$29  &             & RS &          & 2131$-$02  & 4C $-$02.81 & BL &    1.285\\
1430$-$15  & OQ $-$150.2 & Q  &    1.573 & 1648$+$01  &             & RS &          & 2134$+$00  & OX 57       & Q  &    1.932\\
1434$+$23  & OQ 257      & Q  &          & 1652$+$39  & Mark 501    & BL &    0.033 & 2135$-$24  &             & Q  &    0.819\\
1435$-$21  & OQ $-$259   & Q  &    1.187 & 1655$+$07  & OS 92       & Q  &    0.621 & 2136$+$14  & OX 161      & Q  &    2.427\\
1437$-$15  & OQ $-$162   & BL &          & 1656$+$05  & OS 94       & Q  &    0.879 & 2140$-$04  &             & Q  &    0.344\\
1438$+$38  & OQ 363      & Q  &    1.775 & 1656$+$34  & OS 392      & Q  &    1.936 & 2143$-$15  & OX $-$173   & Q  &    0.698\\
1439$+$32  & OQ 366      & Q  &    2.12  & 1657$-$26  &             & RS &          & 2144$+$09  & OX 74       & Q  &    1.113\\
1441$+$25  &             & Q  &    0.062 & 1705$+$01  &             & Q  &    2.576 & 2145$+$06  & DA 562      & Q  &    0.999\\
1442$+$10  & OQ 172      & Q  &    3.535 & 1706$+$00  &             & G  &    0.449 & 2147$+$14  &             & RS &         \\
1443$-$16  & OQ $-$171   & Q  &          & 1706$-$17  & OT $-$111   & RS &          & 2149$+$06  & OX 81       & Q  &    1.364\\
1445$-$16  & OQ $-$176   & Q  &    2.417 & 1717$+$17  & OT 129      & BL &          & 2149$+$05  & OX 82       & Q  &    0.740\\
1449$-$01  & OQ $-$81    & Q  &    1.314 & 1721$+$34  & 4C 34.47    & Q  &    0.206 & 2150$+$17  & OX 183      & BL &         \\
1452$+$30  & OQ 287      & Q  &    0.580 & 1722$+$40  &             & Q  &    1.049 & 2155$-$15  & OX $-$192   & Q  &    0.672\\
1456$+$04  & 4C 04.49    & G  &    0.394 & 1725$+$12  & OT 143.3    & Q  &          & 2200$-$23  &             & Q  &    2.118\\
1502$+$10  & 4C 10.39    & Q  &    1.833 & 1725$+$04  &             & Q  &    0.293 & 2200$+$42  & BL Lacertae & BL &    0.069\\
1502$+$03  &             & G  &    0.413 & 1730$-$13  & NRAO 530    & Q  &    0.902 & 2201$+$31  & 4C 31.63    & Q  &    0.298\\
1504$+$37  & OR 306      & G  &    0.674 & 1732$+$09  & OT 54       & G  &          & 2201$+$17  & OY 101      & Q  &    1.076\\
1504$-$16  & OR $-$102   & Q  &    0.876 & 1741$-$03  & OT $-$68    & Q  &    1.057 & 2201$+$04  & 4C 04.77    & G  &    0.028\\
1508$-$05  & 4C $-$05.64 & Q  &    1.191 & 1743$+$17  & OT 172      & Q  &    1.702 & 2207$+$35  & OY 313      & RS &         \\
1510$-$08  & OR $-$17    & Q  &    0.360 & 1749$+$09  & OT 81       & BL &    0.320 & 2208$-$13  &             & Q  &    0.391\\
1511$-$10  & OR $-$118   & Q  &    1.513 & 1751$+$28  &             & RS &          & 2209$+$08  & DA 574      & Q  &    0.484\\
1511$-$21  & OR $-$218   & G  &    1.179 & 1756$+$23  & OT 295      & Q  &    1.721 & 2209$+$23  &             & Q  &         \\
1514$+$00  & GNZ 25      & G  &    0.052 & 1758$+$38  & OT 398      & Q  &    2.092 & 2214$+$35  & OY 324      & Q  &    0.510\\
1514$+$19  &             & BL &          & 1807$+$27  & 4C 27.41    & Q  &    1.760 & 2215$+$02  &             & Q  &    3.581\\
1514$-$24  & AP Librae   & BL &    0.048 & 1821$+$10  &             & Q  &    1.364 & 2216$-$03  & 4C $-$03.79 & Q  &    0.901\\
1518$+$04  & 4C 04.51    & G  &    1.294 & 1830$+$28  & CTD 108     & Q  &    0.594 & 2223$-$05  & 3C 446      & Q  &    1.404\\
1519$-$27  &             & BL &          & 1848$+$28  &             & Q  &    2.56  & 2223$+$21  & DA 580      & Q  &    1.953\\
1525$+$31  & OR 342      & Q  &    1.380 & 1901$+$31  & 3C 395      & Q  &    0.635 & 2227$-$08  &             & Q  &    1.562\\
1532$+$01  &             & Q  &    1.420 & 1908$-$20  & OV $-$213   & Q  &          & 2229$-$17  & OY $-$150   & Q  &    1.780\\
1535$+$00  &             & Q  &    3.497 & 1908$-$21  & OV $-$214   & RS &          & 2230$+$11  & CTA 102     & Q  &    1.037\\
1538$+$14  & 4C 14.60    & BL &    0.605 & 1920$-$21  & OV $-$235   & RS &          & 2233$-$14  & OY $-$156   & BL & $>$0.609\\
1543$+$00  &             & G  &    0.550 & 1921$-$29  & OV $-$236   & BL &    0.352 & 2234$+$28  & CTD 135     & Q  &    0.795\\
1546$+$02  & OR 78       & Q  &    0.412 & 1923$+$21  & OV 239.7    & RS &          & 2236$+$12  & OY 160      & Q  &         \\
1548$+$05  & 4C 05.64    & Q  &    1.422 & 1936$-$15  & OV $-$161   & Q  &    1.657 & 2239$+$09  &             & Q  &    1.707\\
1548$+$11  & OR 181      & Q  &    0.436 & 1937$-$10  &             & Q  &    3.787 & 2240$-$26  & OY $-$268   & BL &    0.774\\
1550$-$26  &             & Q  &    2.145 & 1947$+$07  & OV 80       & Q  &          & 2243$-$12  & OY $-$172.6 & Q  &    0.630\\
1551$+$13  & OR 186      & Q  &    1.29  & 1958$-$17  & OV $-$198   & Q  &    0.652 & 2245$-$12  &             & Q  &    1.892\\
1555$+$00  &             & Q  &    1.772 & 2002$-$18  & OW $-$105   & Q  &    0.868 & 2245$+$02  &             & Q  &         \\
1555$-$14  &             & G  &    0.097 & 2008$-$15  & OW $-$115   & Q  &    1.180 & 2246$+$20  &             & RS &         \\
1556$-$24  &             & Q  &    2.818 & 2008$-$06  & OW $-$15    & G  &    1.047 & 2247$+$13  & 4C 13.84    & Q  &    0.767\\
1600$+$33  & OS 300      & G  &          & 2012$-$01  &             & BL &          & 2251$+$15  & 3C 454.3    & Q  &    0.859\\
1604$+$31  &             & G  &          & 2029$+$12  & OW 149      & BL &    1.215 & 2251$+$24  & DA 587      & Q  &    2.327\\
1606$+$10  & DA 401      & Q  &    1.226 & 2032$+$10  & OW 154.9    & BL &    0.601 & 2251$+$13  & 4C 13.85    & Q  &    0.677\\
1607$+$26  & OS 211      & G  &    0.473 & 2037$-$25  &             & Q  &    1.574 & 2252$-$09  &             & Q  &    0.606\\
\hline
\end{tabular}
\end{center}
\end{table*}

\clearpage
\begin{table}[t]
{\bf Table 1.} continued
\begin{center}
\begin{tabular}{llll}
\hline
IAU        & Alias       & OI &  $z$    \\
\hline
2253$+$41  & OY 489      & Q  &    1.476\\
2254$+$02  & OY 91.3     & Q  &    2.089\\
2254$+$07  & OY 091      & BL &    0.190\\
2255$+$41  & 4C 41.45    & Q  &    2.15 \\
2255$-$28  &             & Q  &    0.926\\
2256$+$01  &             & Q  &    2.663\\
2300$-$18  & OZ $-$102   & G  &    0.129\\
2303$-$05  & 4C $-$05.95 & Q  &    1.139\\
2307$+$10  & 4C 10.70    & RS &         \\
2318$+$04  & OZ 031      & Q  &    0.623\\
2318$-$19  & OZ $-$130   & G  &         \\
2319$+$31  &             & G  &         \\
2319$+$27  & CTD 139     & Q  &    1.253\\
2320$+$07  & DA 599      & Q  &    2.090\\
2320$-$02  &             & Q  &         \\
2320$-$03  &             & Q  &    1.411\\
2325$-$15  &             & Q  &    2.465\\
2327$+$33  &             & Q  &    1.809\\
2328$+$10  & 4C 10.73    & Q  &    1.489\\
2328$+$31  & OZ 347      & RS &         \\
2329$-$16  & OZ $-$149   & Q  &    1.153\\
2330$+$08  & OZ 50.8     & RS &         \\
2331$-$24  & OZ $-$252   & G  &    0.048\\
2332$-$01  &             & Q  &    1.184\\
2335$-$18  & OZ $-$160   & Q  &    1.450\\
2335$-$02  &             & Q  &    1.072\\
2337$+$26  &             & Q  &         \\
2338$+$33  &             & RS &         \\
2344$+$09  & 4C 09.74    & Q  &    0.673\\
2344$+$092 &             & RS &         \\
2345$-$16  & OZ $-$176   & Q  &    0.576\\
2349$-$01  & 4C $-$01.61 & Q  &    0.174\\
2351$-$00  &             & Q  &    0.463\\
2351$-$15  & OZ $-$187   & Q  &    2.675\\
2354$-$11  &             & Q  &    0.960\\
2355$-$10  &             & Q  &    1.622\\
2356$+$19  & OZ 193      & Q  &    1.066\\
2356$+$38  & OZ 395      & Q  &    2.704\\
\hline
\end{tabular}
\end{center}

\begin{list}{}{}
\item[$^{\mathrm{a}}$]The source was not observed in December, 1997
\item[$^{\mathrm{b}}$]The source was partly resolved at all frequencies
\item[$^{\mathrm{c}}$]We have not registered emission from
this object at any frequency (nothing is present at these coordinates
in NED too)
\end{list}
\end{table}
\clearpage
\setcounter{figure}{3}
\begin{figure*}[p]
\resizebox{\hsize}{!}{\includegraphics[trim=0.1cm 1.7cm 0.6cm 0.3cm]{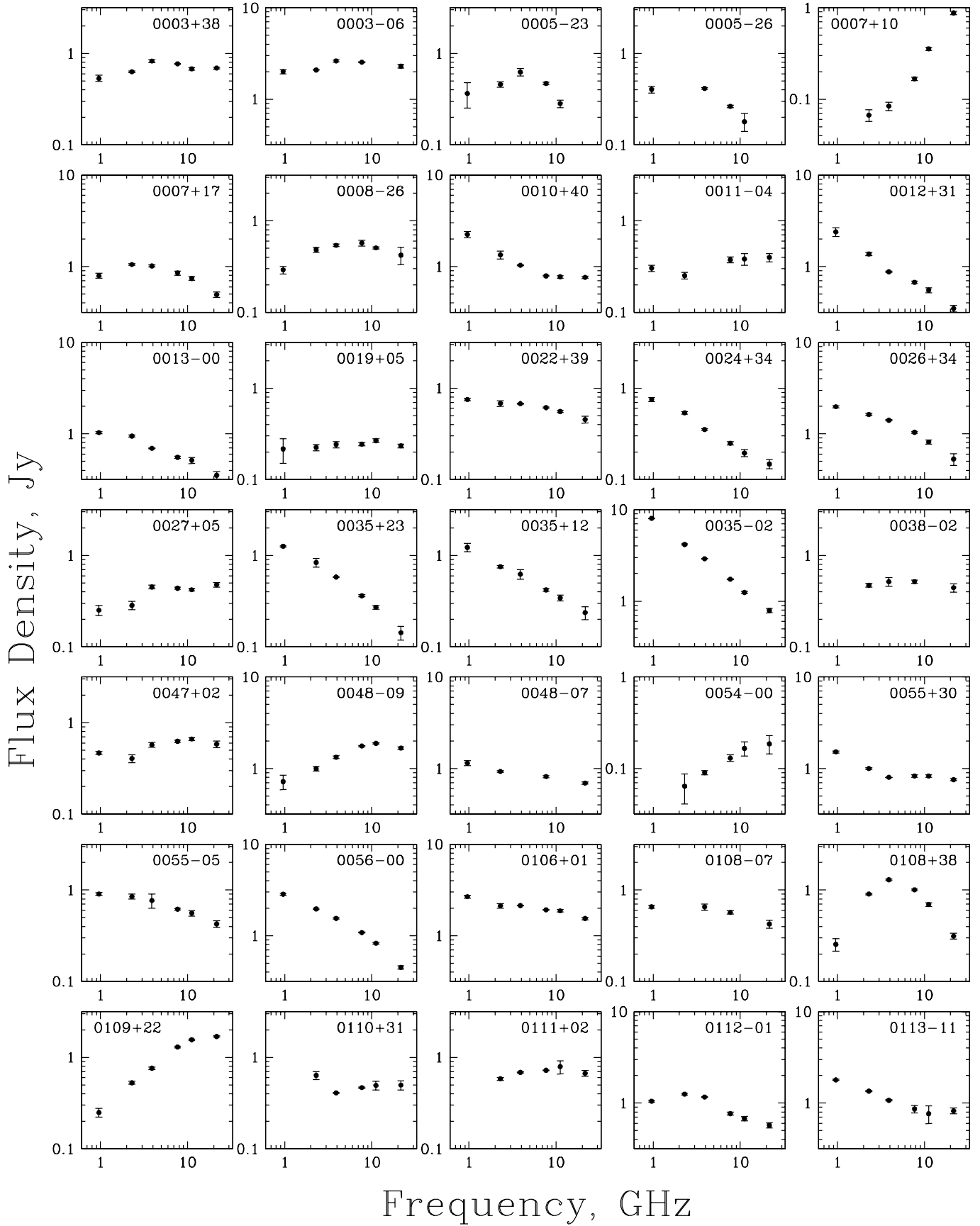}}
\caption{Instantaneous spectra}
\label{spectra_figure}
\end{figure*}

\clearpage
\setcounter{figure}{3}
\begin{figure*}[p]
\resizebox{\hsize}{!}{\includegraphics[trim=0.1cm 1.7cm 0.6cm 0.3cm]{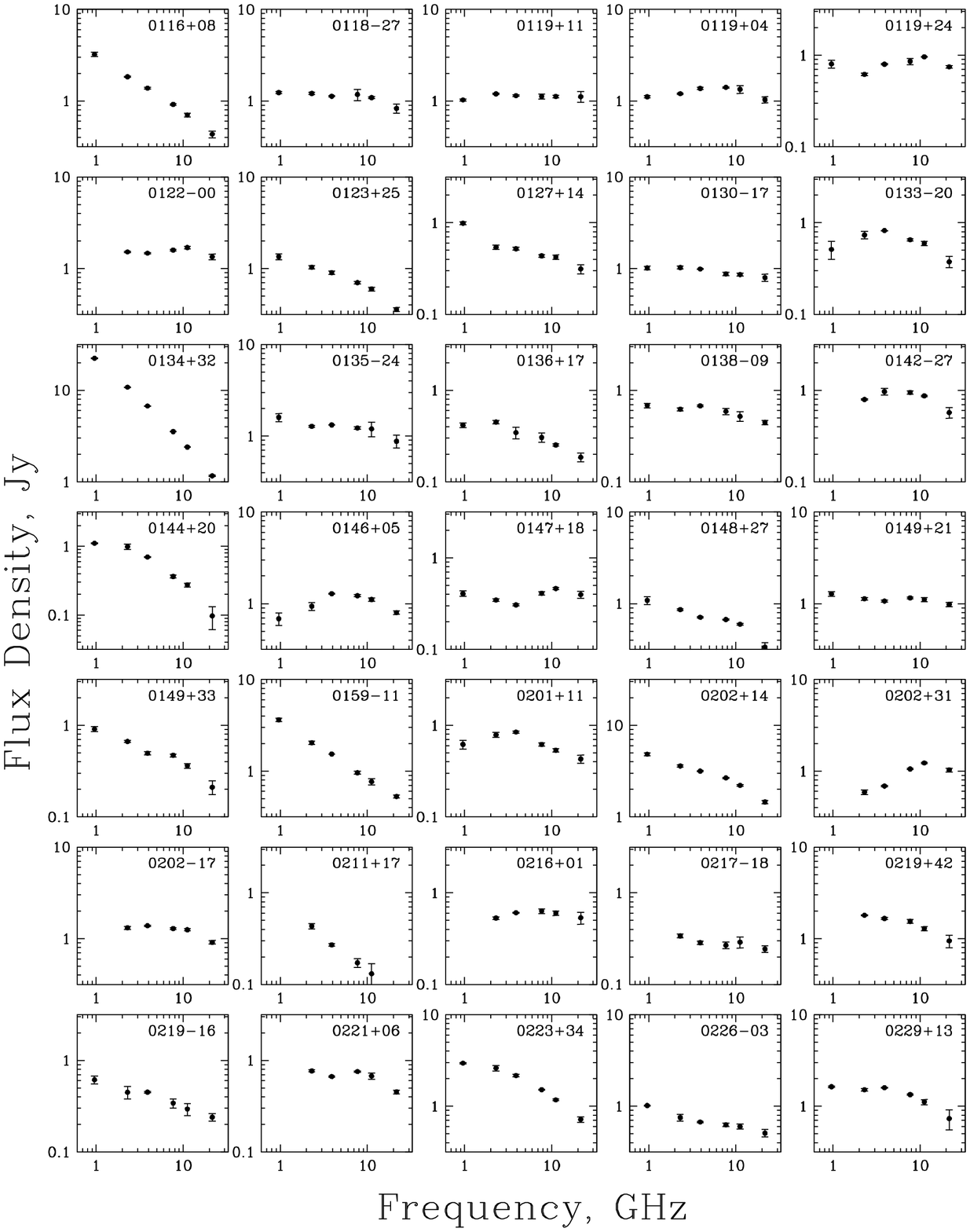}}
\caption{continued}
\end{figure*}

\clearpage
\setcounter{figure}{3}
\begin{figure*}[p]
\resizebox{\hsize}{!}{\includegraphics[trim=0.1cm 1.7cm 0.6cm 0.3cm]{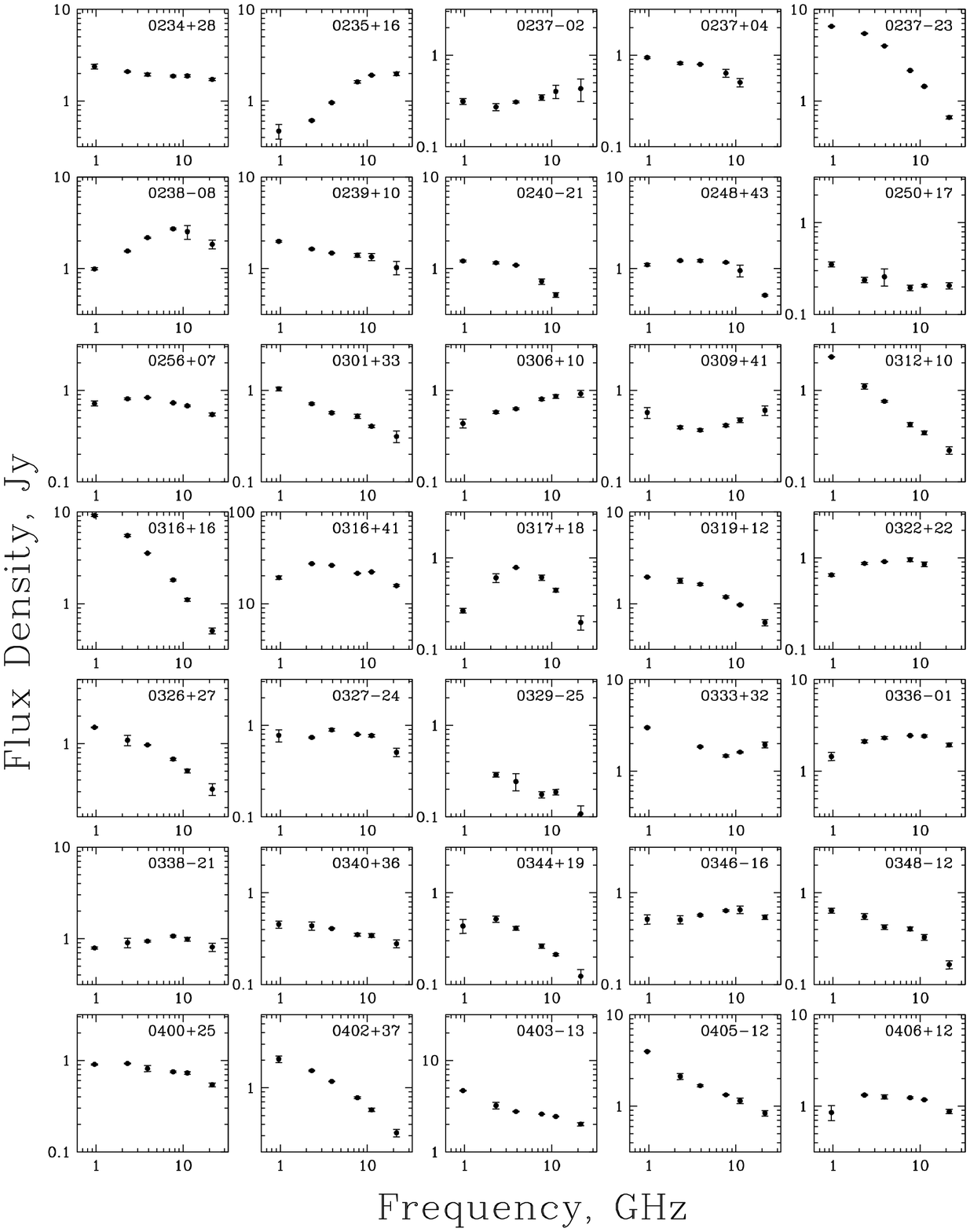}}
\caption{continued}
\end{figure*}

\clearpage
\setcounter{figure}{3}
\begin{figure*}[p]
\resizebox{\hsize}{!}{\includegraphics[trim=0.1cm 1.7cm 0.6cm 0.3cm]{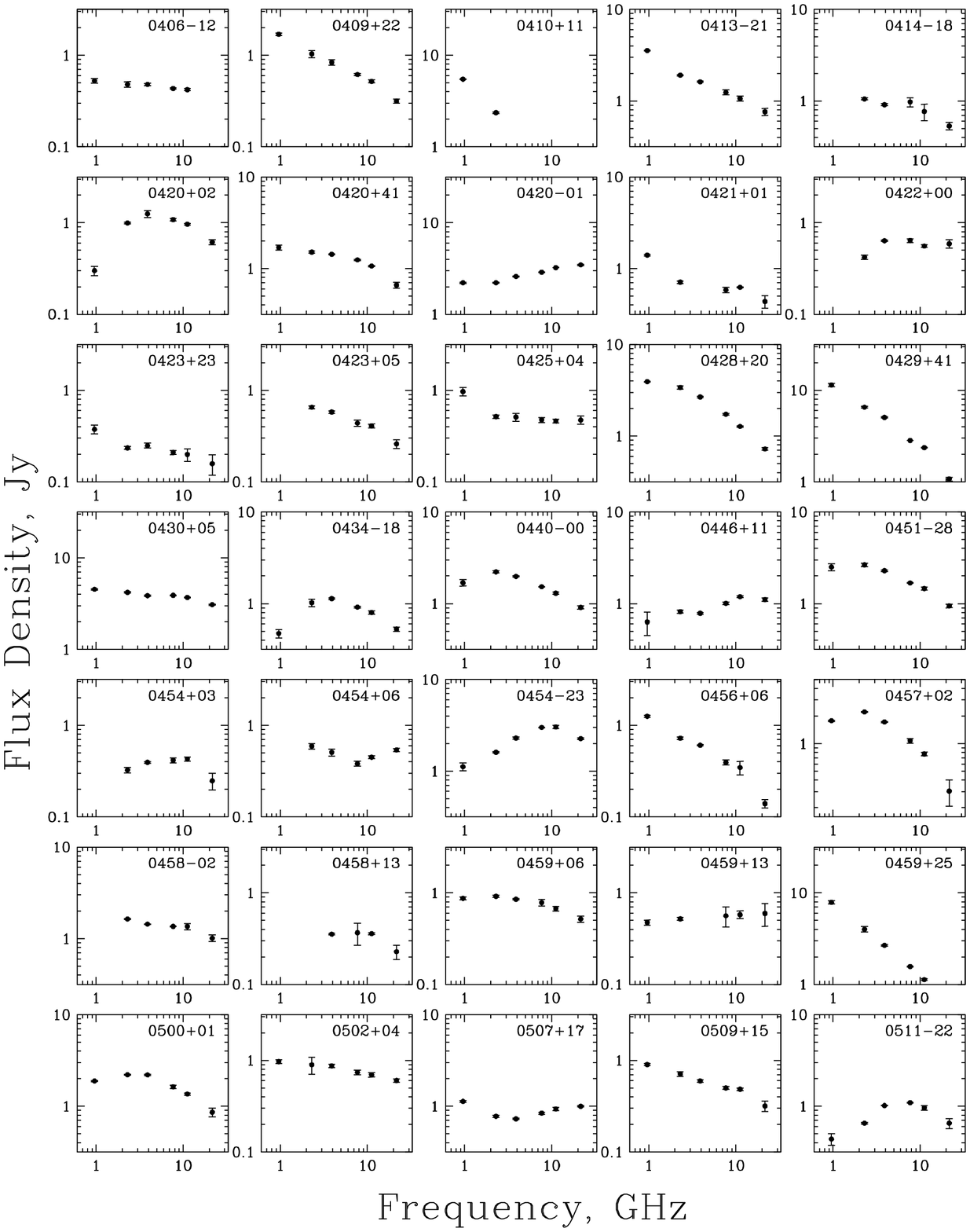}}
\caption{continued}
\end{figure*}

\clearpage
\setcounter{figure}{3}
\begin{figure*}[p]
\resizebox{\hsize}{!}{\includegraphics[trim=0.1cm 1.7cm 0.6cm 0.3cm]{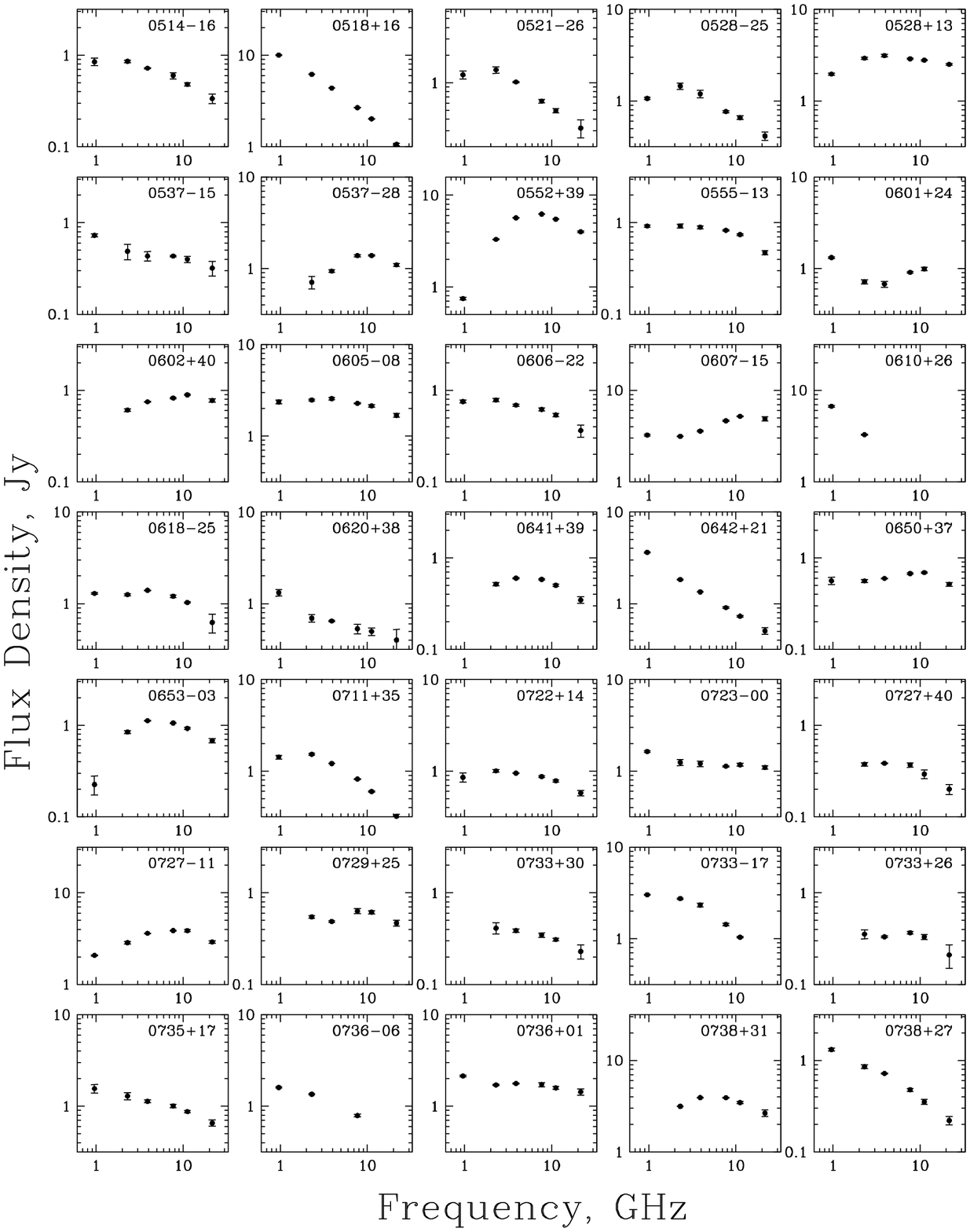}}
\caption{continued}
\end{figure*}

\clearpage
\setcounter{figure}{3}
\begin{figure*}[p]
\resizebox{\hsize}{!}{\includegraphics[trim=0.1cm 1.7cm 0.6cm 0.3cm]{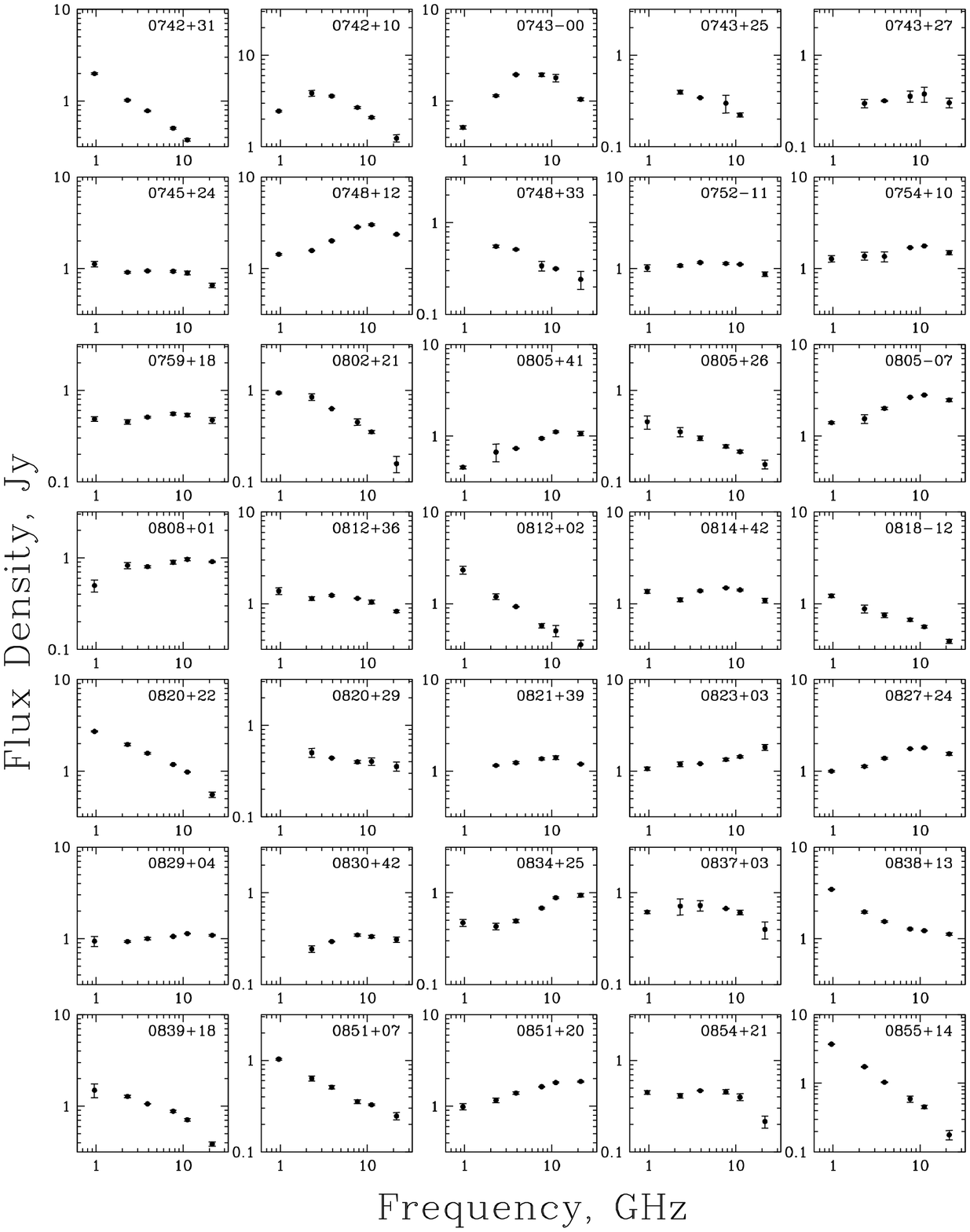}}
\caption{continued}
\end{figure*}

\clearpage
\setcounter{figure}{3}
\begin{figure*}[p]
\resizebox{\hsize}{!}{\includegraphics[trim=0.1cm 1.7cm 0.6cm 0.3cm]{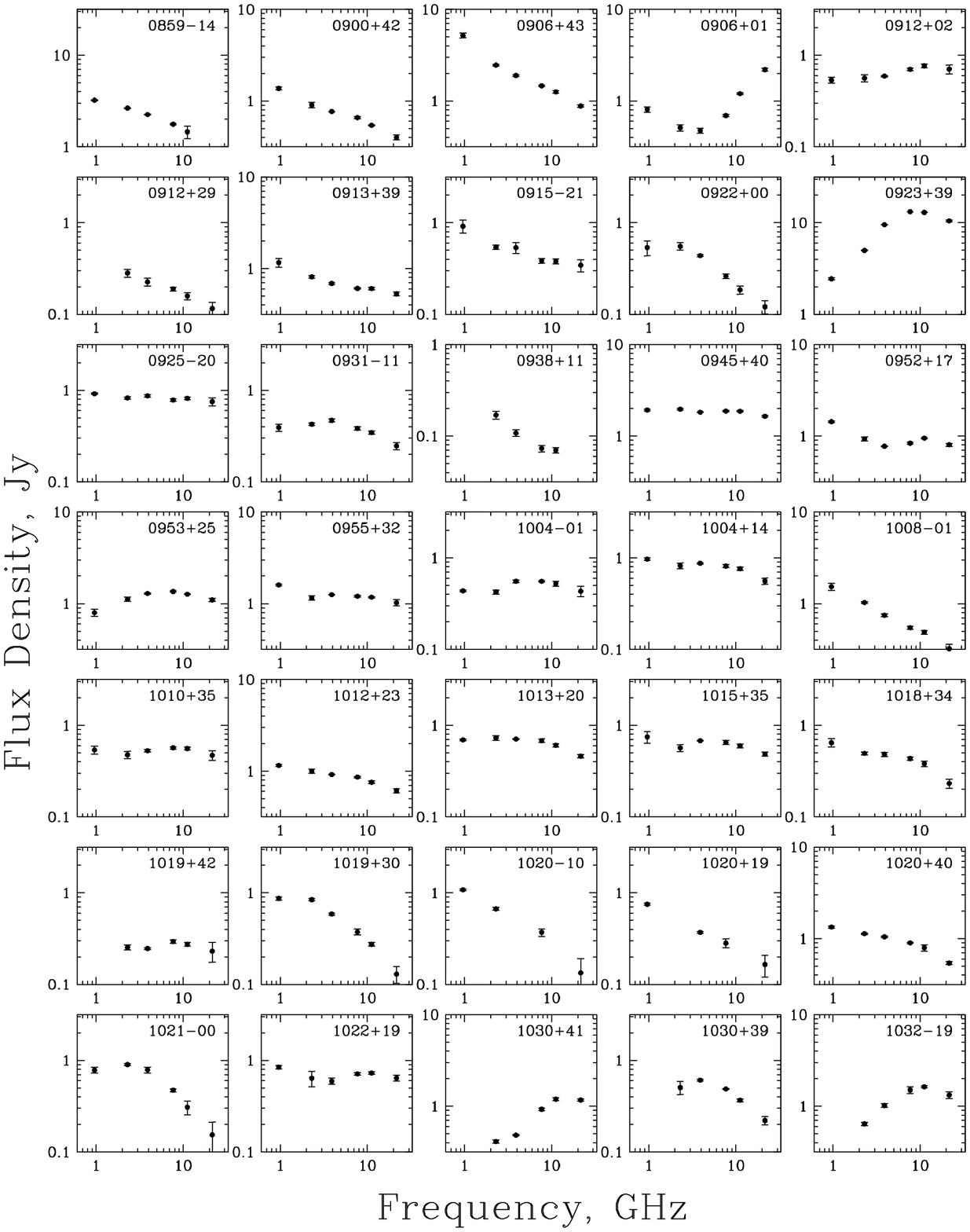}}
\caption{continued}
\end{figure*}

\clearpage
\setcounter{figure}{3}
\begin{figure*}[p]
\resizebox{\hsize}{!}{\includegraphics[trim=0.1cm 1.7cm 0.6cm 0.3cm]{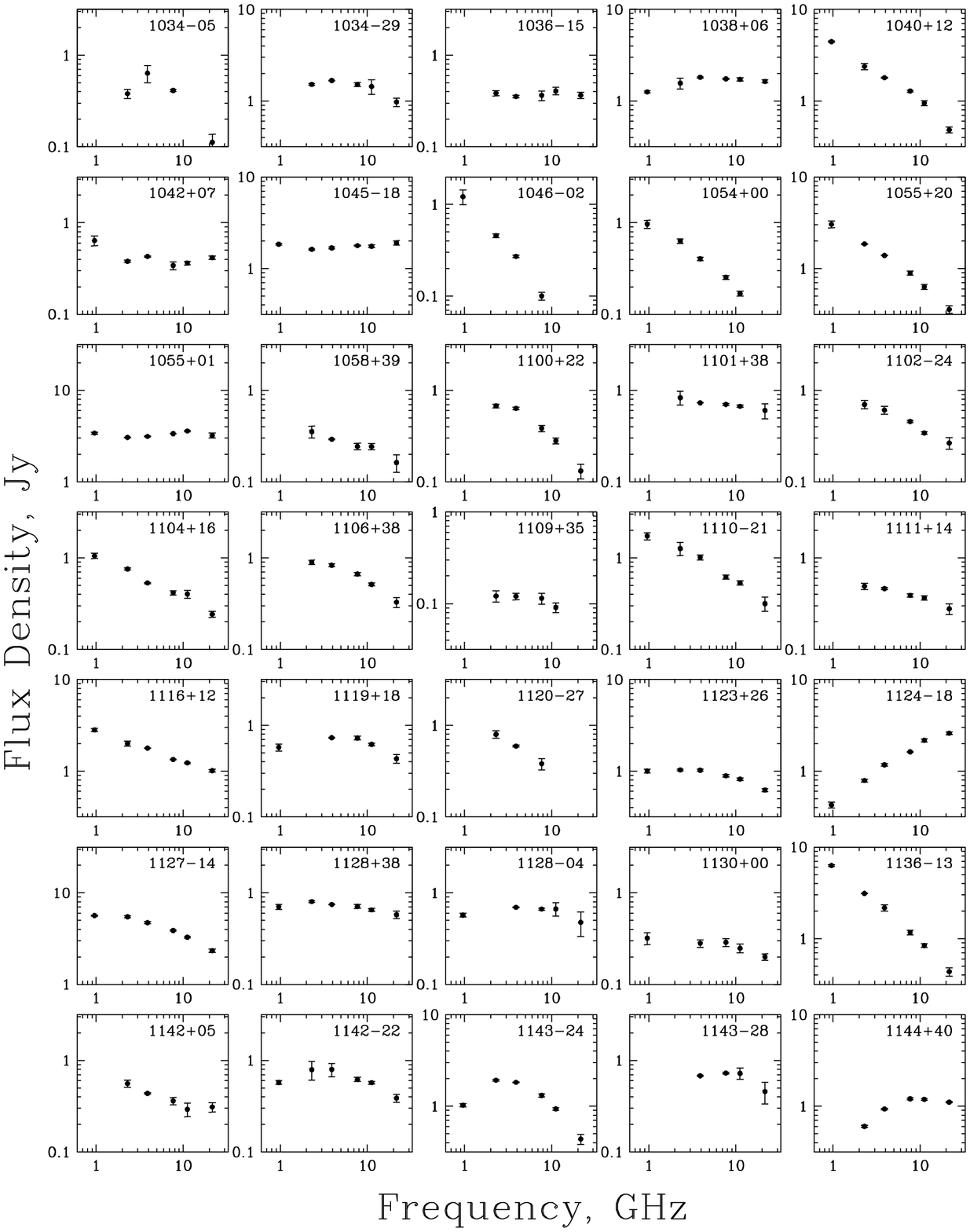}}
\caption{continued}
\end{figure*}

\clearpage
\setcounter{figure}{3}
\begin{figure*}[p]
\resizebox{\hsize}{!}{\includegraphics[trim=0.1cm 1.7cm 0.6cm 0.3cm]{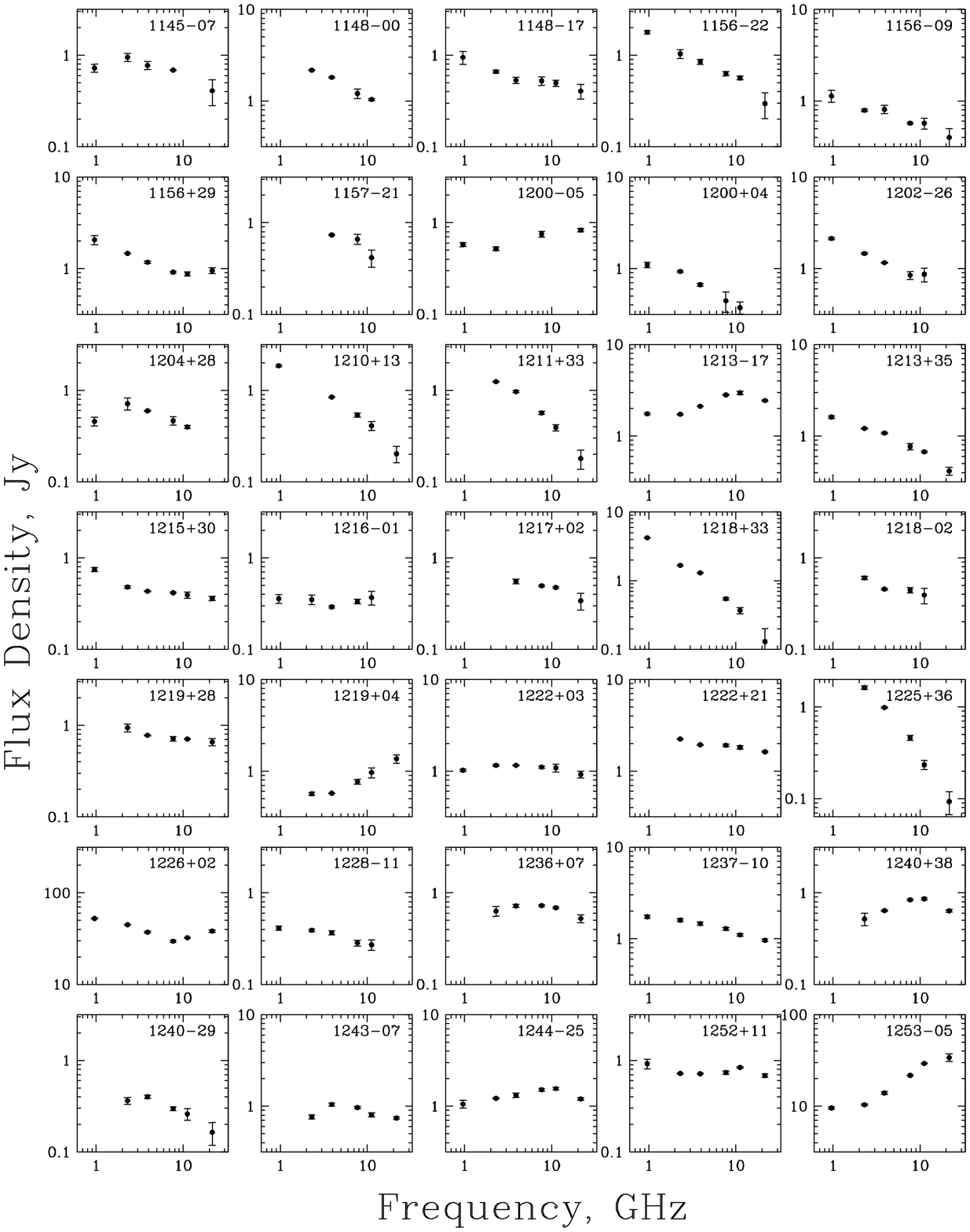}}
\caption{continued}
\end{figure*}

\clearpage
\setcounter{figure}{3}
\begin{figure*}[p]
\resizebox{\hsize}{!}{\includegraphics[trim=0.1cm 1.7cm 0.6cm 0.3cm]{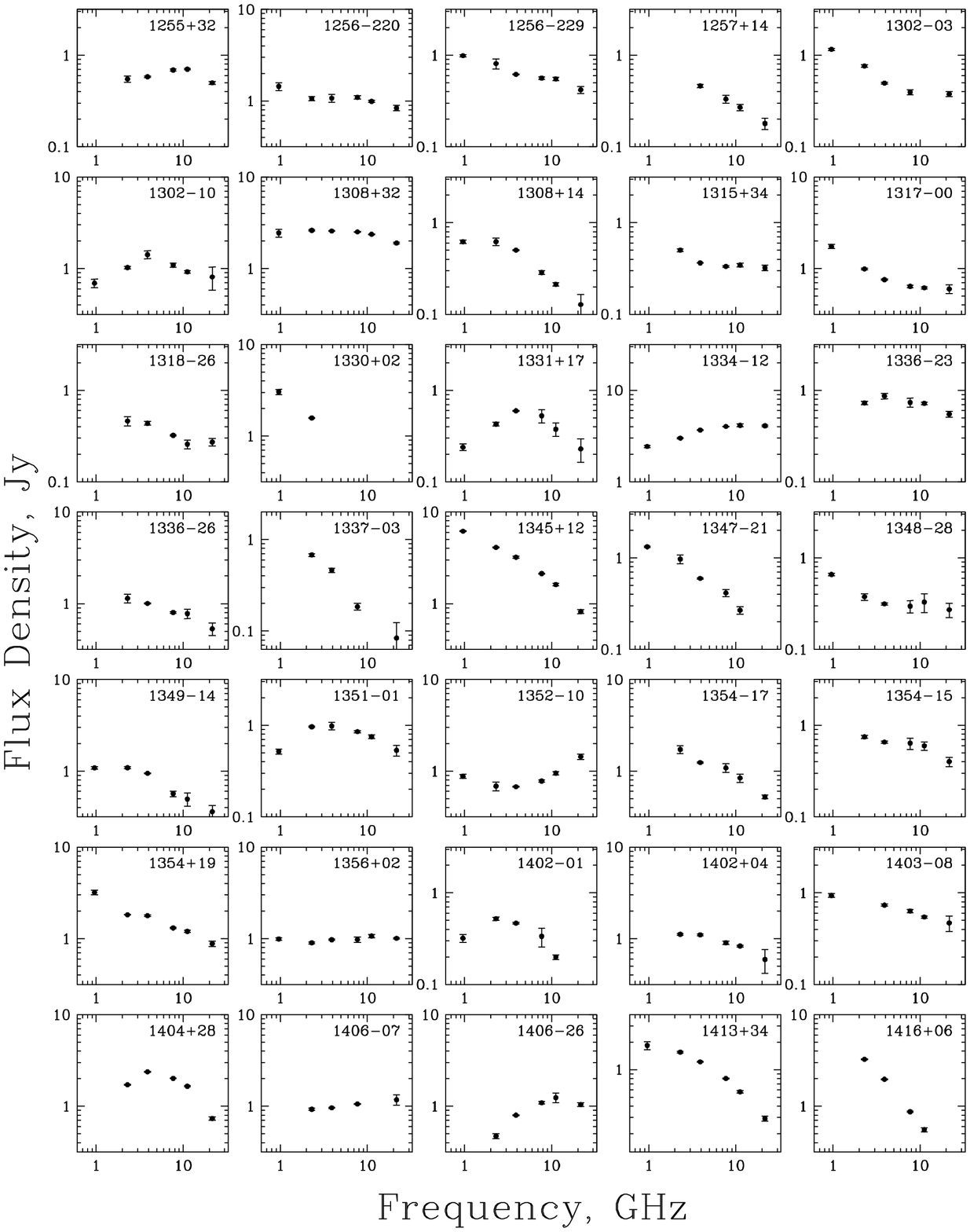}}
\caption{continued}
\end{figure*}

\clearpage
\setcounter{figure}{3}
\begin{figure*}[p]
\resizebox{\hsize}{!}{\includegraphics[trim=0.1cm 1.7cm 0.6cm 0.3cm]{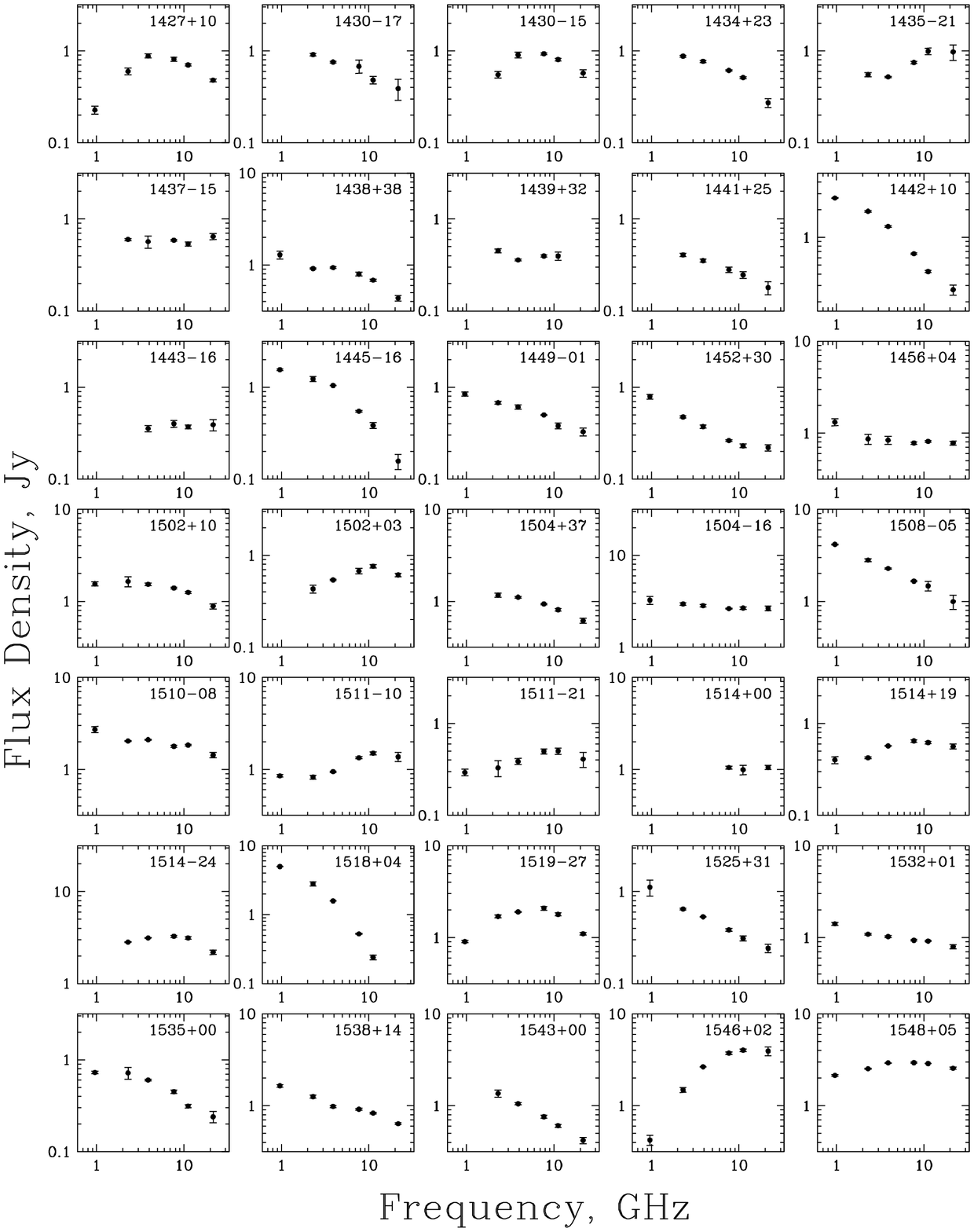}}
\caption{continued}
\end{figure*}

\clearpage
\setcounter{figure}{3}
\begin{figure*}[p]
\resizebox{\hsize}{!}{\includegraphics[trim=0.1cm 1.7cm 0.6cm 0.3cm]{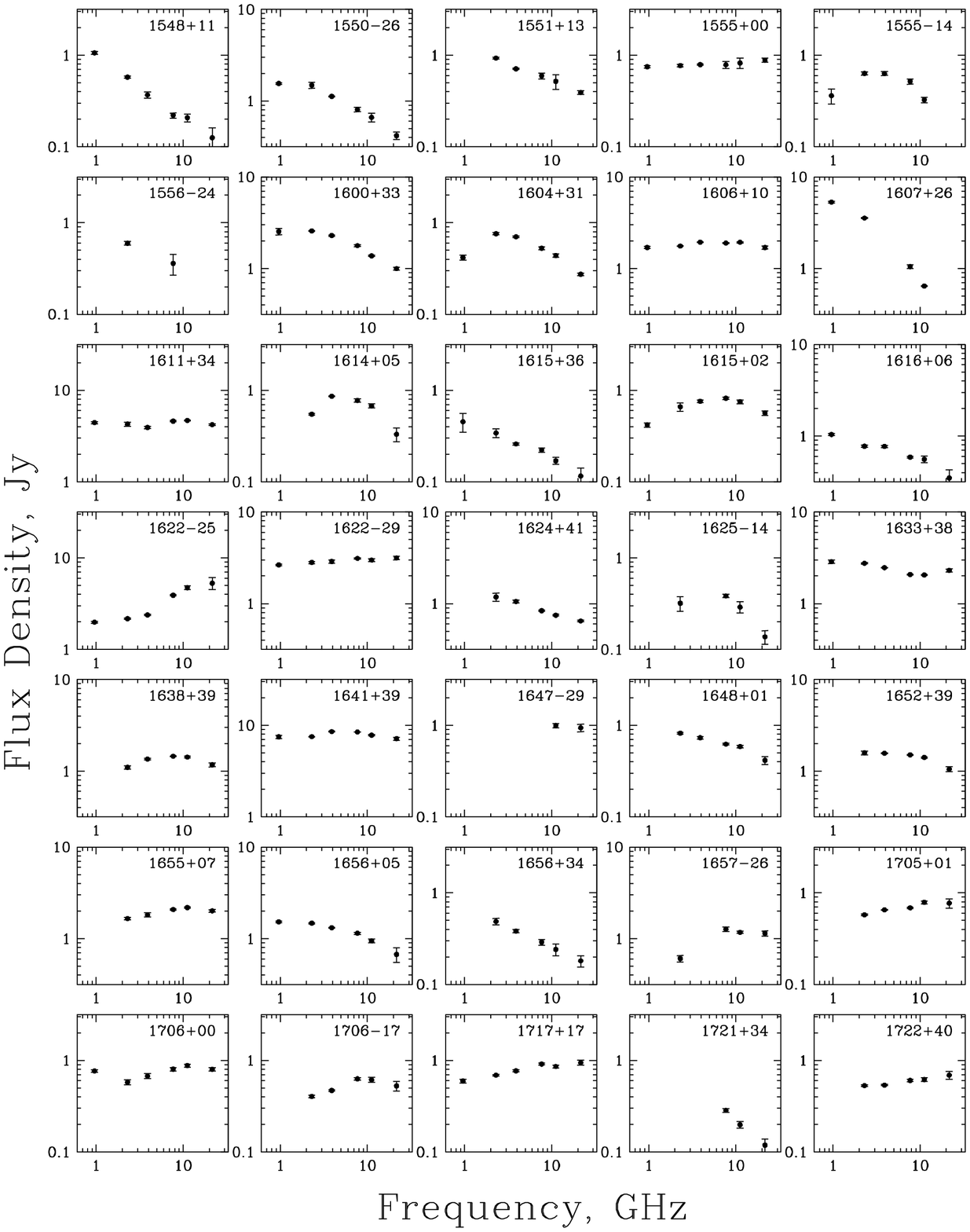}}
\caption{continued}
\end{figure*}

\clearpage
\setcounter{figure}{3}
\begin{figure*}[p]
\resizebox{\hsize}{!}{\includegraphics[trim=0.1cm 1.7cm 0.6cm 0.3cm]{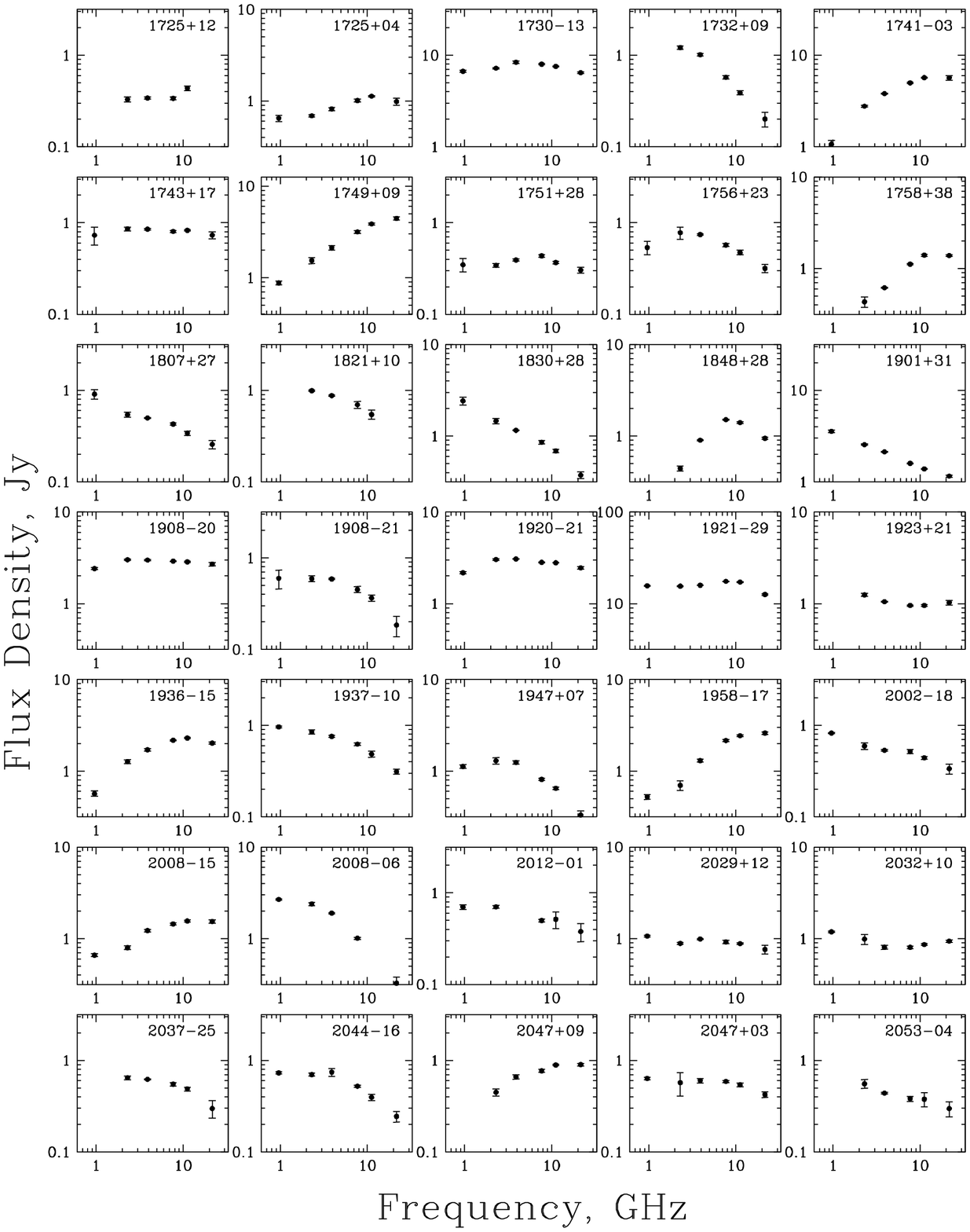}}
\caption{continued}
\end{figure*}

\clearpage
\setcounter{figure}{3}
\begin{figure*}[p]
\resizebox{\hsize}{!}{\includegraphics[trim=0.1cm 1.7cm 0.6cm 0.3cm]{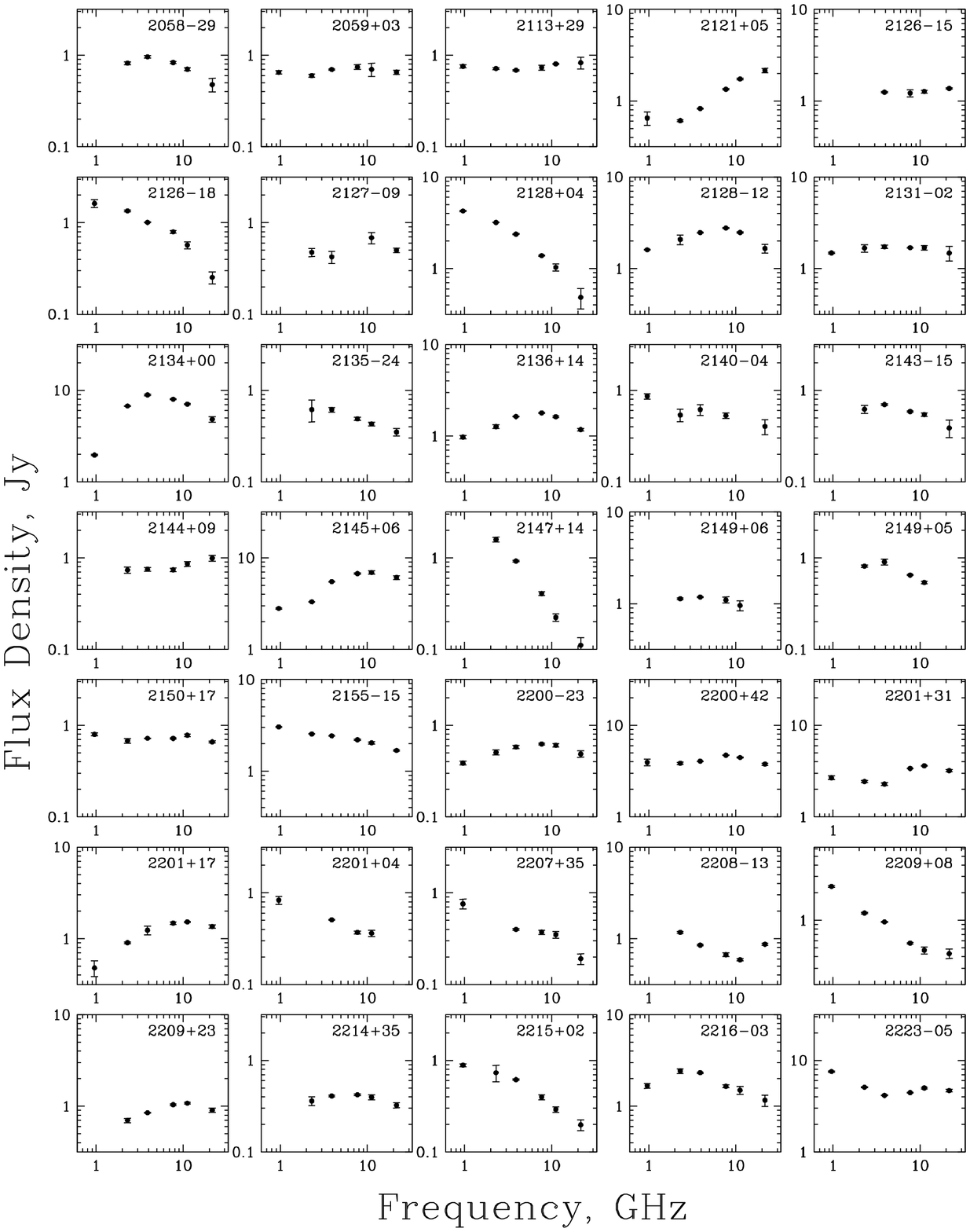}}
\caption{continued}
\end{figure*}

\clearpage
\setcounter{figure}{3}
\begin{figure*}[p]
\resizebox{\hsize}{!}{\includegraphics[trim=0.1cm 1.7cm 0.6cm 0.3cm]{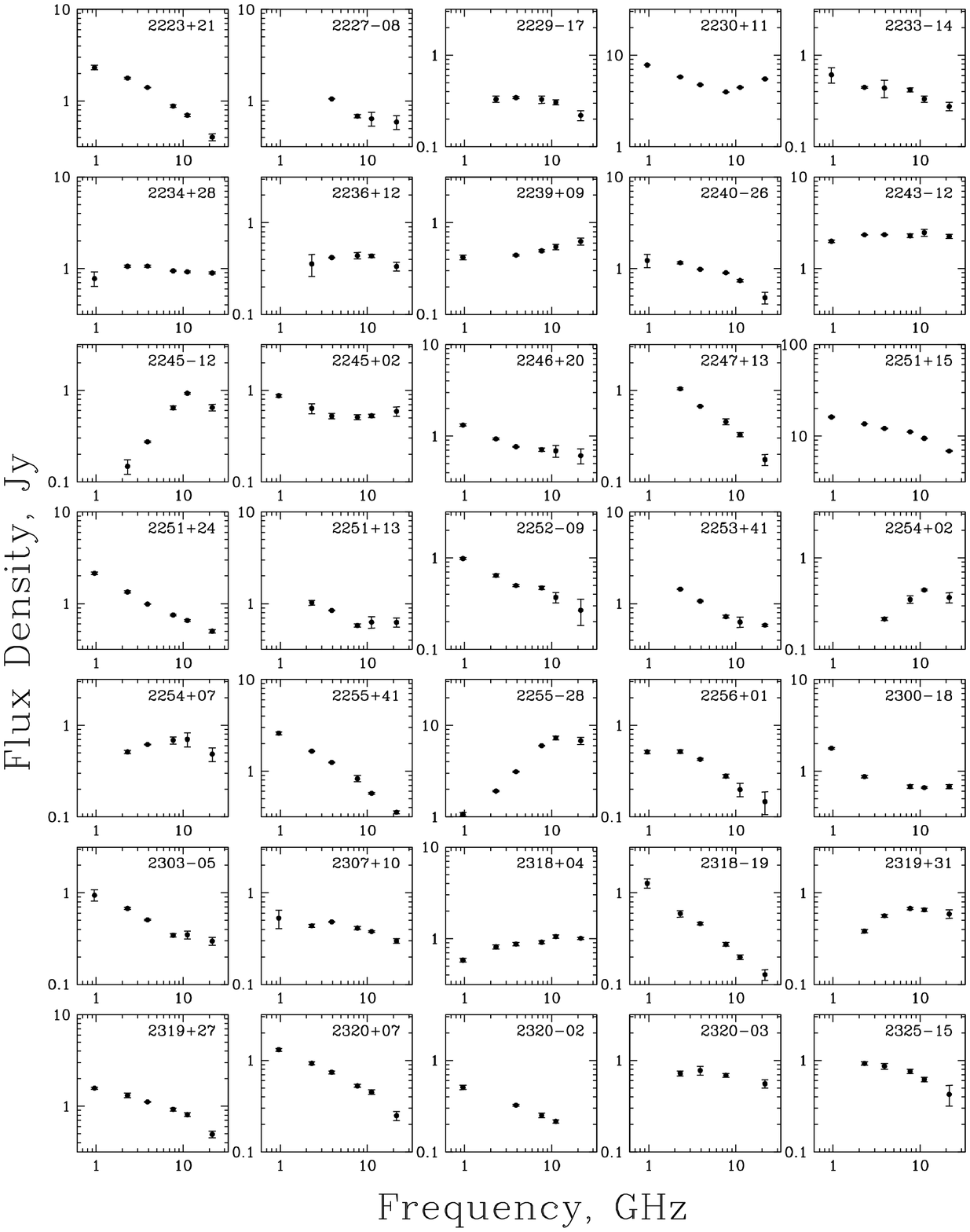}}
\caption{continued}
\end{figure*}

\clearpage
\setcounter{figure}{3}
\begin{figure*}[p]
\resizebox{\hsize}{!}{\includegraphics[trim=0.1cm 8cm 0.6cm 0.3cm]{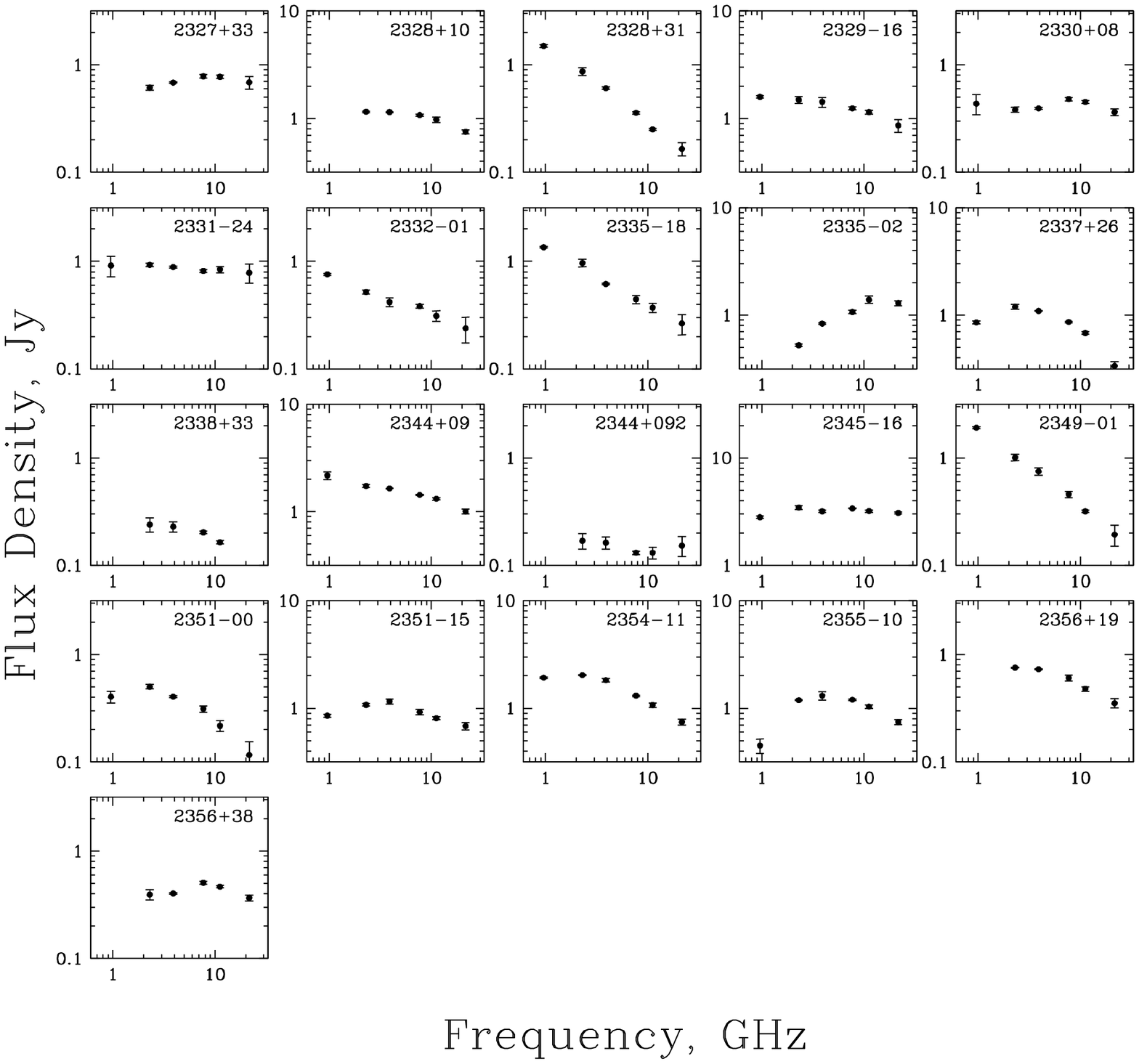}}
\caption{continued}
\end{figure*}


\end{document}